%% file: main.tex
\documentclass[12pt,a4paper]{article}

\usepackage{rotating}

\usepackage{ifthen} 
\newboolean{articletitles}
\setboolean{articletitles}{true} 
\newboolean{uprightparticles}
\setboolean{uprightparticles}{false} 
\newboolean{inbibliography}
\setboolean{inbibliography}{false} 

\input{preamble}
\input{numbers}

\begin{document}
\renewcommand{\thefootnote}{\fnsymbol{footnote}}
\setcounter{footnote}{1}
\input{title}
\renewcommand{\thefootnote}{\arabic{footnote}}
\setcounter{footnote}{0}
\tableofcontents
\cleardoublepage
\pagestyle{plain} 
\setcounter{page}{1}
\pagenumbering{arabic}


\graphicspath{{figs/}}
\allowdisplaybreaks

\input{body}

\setboolean{inbibliography}{true}
\addcontentsline{toc}{section}{References}
\bibliographystyle{LHCb}
\bibliography{references}
\setboolean{inbibliography}{false}

\end{document}

%% file: preamble.tex
\usepackage{microtype}
\usepackage{lineno}  
\usepackage{xspace} 

\usepackage{graphicx}  
\usepackage{color}
\usepackage{colortbl}
\graphicspath{{./figs/}} 

\usepackage{amsmath} 
\usepackage{amssymb}
\usepackage{amsfonts}
\usepackage{mathrsfs,bm,url,color}
\usepackage{upgreek} 
\usepackage{blindtext} 

\newcommand*\patchAmsMathEnvironmentForLineno[1]{%
\expandafter\let\csname old#1\expandafter\endcsname\csname #1\endcsname
\expandafter\let\csname oldend#1\expandafter\endcsname\csname
end#1\endcsname
 \renewenvironment{#1}%
   {\linenomath\csname old#1\endcsname}%
   {\csname oldend#1\endcsname\endlinenomath}%
}
\newcommand*\patchBothAmsMathEnvironmentsForLineno[1]{%
  \patchAmsMathEnvironmentForLineno{#1}%
  \patchAmsMathEnvironmentForLineno{#1*}%
}
\AtBeginDocument{%
\patchBothAmsMathEnvironmentsForLineno{equation}%
\patchBothAmsMathEnvironmentsForLineno{align}%
\patchBothAmsMathEnvironmentsForLineno{flalign}%
\patchBothAmsMathEnvironmentsForLineno{alignat}%
\patchBothAmsMathEnvironmentsForLineno{gather}%
\patchBothAmsMathEnvironmentsForLineno{multline}%
}

\usepackage{hyperref}    
\usepackage[all]{hypcap} 

\input{lhcb-symbols-def} 

\usepackage{cite} 
\usepackage{mciteplus}
\usepackage{overpic}
\usepackage{subfigure}

\usepackage[noabbrev,capitalise]{cleveref}

\usepackage{floatrow}

\newfloatcommand{capbtabbox}{table}[][\FBwidth]

\usepackage{caption}
\usepackage{longtable}
\usepackage{ifpdf}

\newcommand{\re}[2][()] {\ifthenelse{\equal{#1}{()}}{{\ensuremath{{\rm \, Re}}\left(#2\right)}}
                                                    {{\ensuremath{{\rm \, Re}}\left[#2\right]}}}
\newcommand{\im}[2][()] {\ifthenelse{\equal{#1}{()}}{{\ensuremath{{\rm \, Im}}\left(#2\right)}}
                                                    {{\ensuremath{{\rm \, Im}}\left[#2\right]}}}

\definecolor{orange}{rgb}{1,0.5,0}

\def\superkekb  {\mbox{SuperKEKB}\xspace}
\def\besiii  {\mbox{BESIII}\xspace}

\def\Bds     {{\ensuremath{\B_{(\squark)}^0}}\xspace}
\newcommand{\Bsmumu}{\decay{\Bs}{\mup \mun}}

\newcommand{\Bdmumu}{\decay{\Bd}{\mup \mun}}

\newcommand{\Bdsmumu}{\decay{\Bds}{\mup \mun}}
\newcommand{\bsmumu}{\Bsmumu}

\newcommand{\bdmumu}{\Bdmumu} 
\newcommand{\bdsmumu}{\Bdsmumu} 

\def\Bsee     {\decay{\Bs}{\epem}}
\def\Bdee     {\decay{\Bd}{\epem}}

\def\Bstautau     {\decay{\Bs}{\tau^+\tau^-}}
\def\Bdtautau     {\decay{\Bd}{\tau^+\tau^-}}
\newcommand{\BuJpsiK}{\decay{\Bu}{\jpsi \Kp}}

\newcommand{\aunit}[1]{\ensuremath{\text{\,#1}}}       
\def\ab   {\ensuremath{\aunit{ab}}\xspace}
\def\invab   {\ensuremath{\ab^{-1}}\xspace}

\def\bdll     {\decay{\bquark}{\dquark \ell^+ \ell^-}}

\def\BuTopimm {\decay{\Bu}{\pip\mup\mun}}
\def\LbToPpimm{\decay{\Lb}{\proton\pip\mu\mu}}
\def\BsToKstmm{\decay{\Bs}{\Kstarzb\mup\mun}}
\def\BdToKstmm{\decay{\Bd}{\Kstarzb\mup\mun}}
\def\BdToKstee{\decay{\Bd}{\Kstarzb\ep\en}}

\def\bdee     {\decay{\bquark}{\dquark e^+ e^-}}

\def\bsgamma     {\decay{\bquark}{\squark \gamma}}
\def\bdgamma     {\decay{\bquark}{\dquark \gamma}}

\def\BsToPhigamma    {\decay{\Bs}{\phi\gamma}}
\def\LbToLgamma      {\decay{\Lb}{\Lambda \gamma}}
\def\BdToKstgamma    {\decay{\Bd}{\Kstarzb\gamma}}
\def\BuToKpipigamma  {\decay{\Bu}{\Kp\pip\pim\gamma}}

\usepackage{booktabs}
\usepackage{rotating}
\usepackage{multirow}

\newcommand\snowmass{\begin{center}\rule[-0.2in]{\textwidth}{0.01in}\\\rule{\textwidth}{0.01in}\\
\vskip 0.1in Submitted to the Proceedings of the US Community Study\\ 
on the Future of Particle Physics (Snowmass 2021)\\ 
\rule{\textwidth}{0.01in}\\\rule[+0.2in]{\textwidth}{0.01in} \end{center}}

%% file: lhcb-symbols-def.tex
\def\lhcb {\mbox{LHCb}\xspace}
\def\ux85 {\mbox{UX85}\xspace}

\def\atlas  {\mbox{ATLAS}\xspace}
\def\cms    {\mbox{CMS}\xspace}
\def\babar  {\mbox{BaBar}\xspace}
\def\belle  {\mbox{Belle}\xspace}
\def\belleii  {\mbox{BelleII}\xspace}

\ifthenelse{\boolean{uprightparticles}}%
{

 \def\Pmu         {\ensuremath{\upmu}\xspace}

 \def\Ppi         {\ensuremath{\uppi}\xspace}

 \def\Ppsi        {\ensuremath{\uppsi}\xspace}

 \def\PDelta      {\ensuremath{\Delta}\xspace}                 
 \def\PXi      {\ensuremath{\Xi}\xspace}                 
 \def\PLambda      {\ensuremath{\Lambda}\xspace}                 
 \def\PSigma      {\ensuremath{\Sigma}\xspace}                 
 \def\POmega      {\ensuremath{\Omega}\xspace}                 
 \def\PUpsilon      {\ensuremath{\Upsilon}\xspace}                 
 
 \def\PB      {\ensuremath{\mathrm{B}}\xspace}                 
                  
 \def\PD      {\ensuremath{\mathrm{D}}\xspace}

 \def\PJ      {\ensuremath{\mathrm{J}}\xspace}                 
 \def\PK      {\ensuremath{\mathrm{K}}\xspace}

 \def\Pb      {\ensuremath{\mathrm{b}}\xspace}                 
 \def\Pc      {\ensuremath{\mathrm{c}}\xspace}                 
 \def\Pd      {\ensuremath{\mathrm{d}}\xspace}                 
 \def\Pe      {\ensuremath{\mathrm{e}}\xspace}

 \def\Pi      {\ensuremath{\mathrm{i}}\xspace}

 \def\Pp      {\ensuremath{\mathrm{p}}\xspace}                 
 \def\Pq      {\ensuremath{\mathrm{q}}\xspace}                 
                  
 \def\Ps      {\ensuremath{\mathrm{s}}\xspace}                 
                  
 \def\Pu      {\ensuremath{\mathrm{u}}\xspace}

}
{

 \def\Pmu         {\ensuremath{\mu}\xspace}

 \def\Ppi         {\ensuremath{\pi}\xspace}

 \def\Ppsi        {\ensuremath{\psi}\xspace}                 
                  
 \mathchardef\PDelta="7101
 \mathchardef\PXi="7104
 \mathchardef\PLambda="7103
 \mathchardef\PSigma="7106
 \mathchardef\POmega="710A
 \mathchardef\PUpsilon="7107
                  
 \def\PB      {\ensuremath{B}\xspace}                 
                  
 \def\PD      {\ensuremath{D}\xspace}

 \def\PJ      {\ensuremath{J}\xspace}                 
 \def\PK      {\ensuremath{K}\xspace}

 \def\Pb      {\ensuremath{b}\xspace}                 
 \def\Pc      {\ensuremath{c}\xspace}                 
 \def\Pd      {\ensuremath{d}\xspace}                 
 \def\Pe      {\ensuremath{e}\xspace}

 \def\Pi      {\ensuremath{i}\xspace}

 \def\Pp      {\ensuremath{p}\xspace}                 
 \def\Pq      {\ensuremath{q}\xspace}                 
                  
 \def\Ps      {\ensuremath{s}\xspace}                 
                  
 \def\Pu      {\ensuremath{u}\xspace}

}

\def\en         {\ensuremath{\Pe^-}\xspace}   
\def\ep         {\ensuremath{\Pe^+}\xspace}
 
\def\epem       {\ensuremath{\Pe^+\Pe^-}\xspace}

\def\mup        {\ensuremath{\Pmu^+}\xspace}
\def\mun        {\ensuremath{\Pmu^-}\xspace} 

\def\quark     {\ensuremath{\Pq}\xspace}

\def\uquark    {\ensuremath{\Pu}\xspace}

\def\dquark    {\ensuremath{\Pd}\xspace}

\def\squark    {\ensuremath{\Ps}\xspace}

\def\cquark    {\ensuremath{\Pc}\xspace}

\def\bquark    {\ensuremath{\Pb}\xspace}

\def\pion  {\ensuremath{\Ppi}\xspace}
\def\piz   {\ensuremath{\pion^0}\xspace}

\def\pip   {\ensuremath{\pion^+}\xspace}
\def\pim   {\ensuremath{\pion^-}\xspace}

\def\kaon  {\ensuremath{\PK}\xspace}
  \def\Kbar  {\kern 0.2em\overline{\kern -0.2em \PK}{}\xspace}

\def\Kz    {\ensuremath{\kaon^0}\xspace}
\def\Kzb   {\ensuremath{\Kbar^0}\xspace}
\def\KzKzb {\ensuremath{\Kz \kern -0.16em \Kzb}\xspace}
\def\Kp    {\ensuremath{\kaon^+}\xspace}
\def\Km    {\ensuremath{\kaon^-}\xspace}

\def\KpKm  {\ensuremath{\Kp \kern -0.16em \Km}\xspace}

\def\Kstarzb {\ensuremath{\Kbar^{*0}}\xspace}

\def\Dbar    {\kern 0.2em\overline{\kern -0.2em \PD}{}\xspace}
\def\D       {\ensuremath{\PD}\xspace}

\def\Dz      {\ensuremath{\D^0}\xspace}
\def\Dzb     {\ensuremath{\Dbar^0}\xspace}
\def\DzDzb   {\ensuremath{\Dz {\kern -0.16em \Dzb}}\xspace}
\def\Dp      {\ensuremath{\D^+}\xspace}
\def\Dm      {\ensuremath{\D^-}\xspace}

\def\DpDm    {\ensuremath{\Dp {\kern -0.16em \Dm}}\xspace}

\def\B       {\ensuremath{\PB}\xspace}
\def\Bbar    {\ensuremath{\kern 0.18em\overline{\kern -0.18em \PB}{}}\xspace}

\def\Bz      {\ensuremath{\B^0}\xspace}
\def\Bzb     {\ensuremath{\Bbar^0}\xspace}
\def\Bu      {\ensuremath{\B^+}\xspace}

\def\Bpm     {\ensuremath{\B^\pm}\xspace}

\def\Bd      {\ensuremath{\B^0}\xspace}
\def\Bs      {\ensuremath{\B^0_\squark}\xspace}
\def\Bsb     {\ensuremath{\Bbar^0_\squark}\xspace}

\def\Bcpm    {\ensuremath{\B_\cquark^\pm}\xspace}

\def\jpsi     {\ensuremath{{\PJ\mskip -3mu/\mskip -2mu\Ppsi\mskip 2mu}}\xspace}

  \def\Y#1S{\ensuremath{\PUpsilon{(#1S)}}\xspace}

\def\FourS {\Y4S}
\def\FiveS {\Y5S}

\def\proton      {\ensuremath{\Pp}\xspace}

\def\L {\ensuremath{\PLambda}\xspace}
\def\Lbar {\ensuremath{\kern 0.1em\overline{\kern -0.1em\PLambda}}\xspace}

\def\Lb      {\ensuremath{\L^0_\bquark}\xspace}
\def\Lbbar   {\ensuremath{\Lbar^0_\bquark}\xspace}

\newcommand{\BRof}[1]{\ensuremath{{\cal B}(#1)}\xspace}
\newcommand{\decay}[2]{\ensuremath{#1\!\to #2}\xspace}         

\def\to                 {\ensuremath{\rightarrow}\xspace}

\def\CP                {\ensuremath{C\!P}\xspace}

\def\BdToKstmm    {\decay{\Bd}{\Kstarz\mup\mun}}

\def\BsToPhimm    {\decay{\Bs}{\phi\mup\mun}}

\def\bsll     {\decay{\bquark}{\squark \ell^+ \ell^-}}

\def\AT#1     {\ensuremath{A_{\mathrm{T}}^{#1}}\xspace}           

\def\Bsmm     {\decay{\Bs}{\mup\mun}}

\def\C#1      {\ensuremath{\mathcal{C}_{#1}}\xspace}                       
\def\Cp#1     {\ensuremath{\mathcal{C}_{#1}^{'}}\xspace}                    
\def\Ceff#1   {\ensuremath{\mathcal{C}_{#1}^{\mathrm{(eff)}}}\xspace}        
\def\Cpeff#1  {\ensuremath{\mathcal{C}_{#1}^{'\mathrm{(eff)}}}\xspace}       
\def\Ope#1    {\ensuremath{\mathcal{O}_{#1}}\xspace}                       
\def\Opep#1   {\ensuremath{\mathcal{O}_{#1}^{'}}\xspace}                    

\newcommand{\tev}{\ensuremath{\mathrm{\,Te\kern -0.1em V}}\xspace}
\newcommand{\gev}{\ensuremath{\mathrm{\,Ge\kern -0.1em V}}\xspace}
\newcommand{\mev}{\ensuremath{\mathrm{\,Me\kern -0.1em V}}\xspace}
\newcommand{\kev}{\ensuremath{\mathrm{\,ke\kern -0.1em V}}\xspace}
\newcommand{\ev}{\ensuremath{\mathrm{\,e\kern -0.1em V}}\xspace}
\newcommand{\gevc}{\ensuremath{{\mathrm{\,Ge\kern -0.1em V\!/}c}}\xspace}
\newcommand{\mevc}{\ensuremath{{\mathrm{\,Me\kern -0.1em V\!/}c}}\xspace}
\newcommand{\gevcc}{\ensuremath{{\mathrm{\,Ge\kern -0.1em V\!/}c^2}}\xspace}
\newcommand{\gevgevcccc}{\ensuremath{{\mathrm{\,Ge\kern -0.1em V^2\!/}c^4}}\xspace}
\newcommand{\mevcc}{\ensuremath{{\mathrm{\,Me\kern -0.1em V\!/}c^2}}\xspace}

\def\mum  {\ensuremath{\,\upmu\rm m}\xspace}

\def\mub{\ensuremath{\rm \,\upmu b}\xspace}

\def\nb {\ensuremath{\rm \,nb}\xspace}

\def\invfb   {\ensuremath{\mbox{\,fb}^{-1}}\xspace}

\def\ps   {\ensuremath{{\rm \,ps}}\xspace}

\def\gsim{{~\raise.15em\hbox{$>$}\kern-.85em
          \lower.35em\hbox{$\sim$}~}\xspace}
\def\lsim{{~\raise.15em\hbox{$<$}\kern-.85em
          \lower.35em\hbox{$\sim$}~}\xspace}

\def\tell1  {TELL1\xspace}
\def\ukl1   {UKL1\xspace}



%% file: numbers.tex
\def\ATLASBsmumuMeasure{\ensuremath{\left( 2.8 \,^{+0.8}_{-0.7} \right) \times 10^{-9}}\xspace}
\def\ATLASBdmumuMeasure{\ensuremath{\left( -1.9 \pm 1.6 \right) \times 10^{-10}}\xspace}
\def\ATLASBsmumuSignificance{\ensuremath{4.6}\xspace}
\def\ATLASBdmumuUpperLimit{\ensuremath{2.1 \times 10^{-10}}\xspace}

\def\CMSBsmumuMeasure{\ensuremath{\left[ 2.9 \,^{+0.7}_{-0.6}\rm{(exp)} \pm 0.2\rm{(frag)} \right] \times 10^{-9} \xspace}}
\def\CMSBsmumuSignificance{\ensuremath{5.6}\xspace}
\def\CMSBdmumuMeasure{\ensuremath{\left(0.8 \,^{+1.4}_{-1.3} \right) \times 10^{-10}}\xspace}
\def\CMSBdmumuSignificance{\ensuremath{1.0}\xspace}
\def\CMSBdmumuUpperLimit{\ensuremath{3.6 \times 10^{-10}}\xspace}

\def\Bdobslimitnf{\ensuremath{2.6\times 10^{-10}}\xspace} 

\def\Bsbr{\ensuremath{\left(3.09^{\,+\,0.46\,+\,0.15}_{\,-\,0.43\,-\,0.11}\right)\times 10^{-9}}\xspace}
\def\Bdbr{\ensuremath{\left(1.2^{\,+\,0.8}_{\,-\,0.7}\pm 0.1\right)\times 10^{-10}}\xspace}

\def\RmumuLHCb{\ensuremath{0.039^{\,+\,0.030\,+\,0.006}_{\,-\,0.024\,-\,0.004}}\xspace}

\def\Bseelimit{\ensuremath{1.1\times 10^{-8}}\xspace} 
\def\Bdeelimit{\ensuremath{3.0\times 10^{-9}}\xspace} 

\def\Bstautaulimit{\ensuremath{6.8\times 10^{-3}}\xspace} 
\def\Bdtautaulimit{\ensuremath{2.1\times 10^{-3}}\xspace} 

\def\BsmumuCurrent{\ensuremath{\pm 0.46}}
\def\BsmumuLHCbUI{\ensuremath{\pm 0.30 }}
\def\BsmumuLHCbUII{\ensuremath{\pm 0.16}}
\def\BsmumuATLASHL{\ensuremath{\pm (0.50)}}
\def\BsmumuCMSHL{\ensuremath{\pm 0.39 }}

\def\RmumuCurrent{\ensuremath{70\%}}
\def\RmumuLHCbUI{\ensuremath{34\%}}
\def\RmumuLHCbUII{\ensuremath{10\%}}
\def\RmumuATLASHL{\ensuremath{-}}
\def\RmumuCMSHL{\ensuremath{21\%}}

\def\tauCurrent{\ensuremath{14\%}}
\def\taumumuLHCbUI{\ensuremath{\pm 0.16\ps}}
\def\taumumuLHCbUII{\ensuremath{\pm 0.04\ps}}
\def\taumumuATLASHL{\ensuremath{-}}
\def\taumumuCMSHL{\ensuremath{\pm 0.05\ps}}

%% file: title.tex
\begin{titlepage}

\snowmass
\vspace*{1.5cm}

{\bf\boldmath\huge
\begin{center}
Rare decays of \bquark and \cquark hadrons
\end{center}
}

\vspace*{0.5cm}

\begin{center}
F.~Archilli$^{1}$,
W.~Altmannshofer$^{2}$
\bigskip\\
{\it\footnotesize
$^1$Physikalisches Institut, Ruprecht-Karls-Universit{\"a}t Heidelberg, Heidelberg, Germany\\
$^2$Department of Physics, University of California Santa Cruz, and Santa Cruz Institute for Particle Physics, 1156 High St., Santa Cruz, CA 95064, USA}
\end{center}

\vspace{\fill}

\begin{abstract}
\input{abstract}
\end{abstract}

\vspace{\fill}

\end{titlepage}

\pagestyle{empty}  


%% file: abstract.tex
In this white paper for the Snowmass process, we review the status and prospects of the field of rare decays of $b$ and $c$ hadrons. The role that rare decays play in the search for physics beyond the Standard Model is emphasised. We stress the complementarity of a large set of relevant processes and outline the most promising directions. The experimental opportunities at Belle II, BES III, ATLAS, CMS, LHCb, and at future machines are discussed. We also summarize the challenges that need to be addressed on the theory side to achieve theory uncertainties for rare decays that match the expected experimental sensitivities. 

%% file: body.tex
\section{Executive summary} \label{sec:exec}

Flavor changing neutral current (FCNC) processes, rare decays of $b$ and $c$ flavored hadrons in particular, are widely recognized as important probes of physics beyond the Standard Model (SM)~\cite{Blake:2016olu,Gisbert:2020vjx}. The absence of direct signs of new particles at the LHC strengthens their role as tools for indirect discovery of new physics at the TeV scale and beyond.

Rare $b$ and $c$ decays are strongly suppressed in the SM and therefore potentially sensitive to very high new physics scales of several 10's to 100 TeV. Fully exploiting this potential requires precise experimental measurements and equally precise SM predictions. 
The existing experiments, \belleii, \besiii, \atlas, \cms, and \lhcb, as well as future machines, in particular $e^+e^-$ colliders running on the $Z$ pole, provide ample opportunities to measure many rare decays with unprecedented precision. On the theory side, further improvement in the description of the hadronic physics that governs the rare decays is required. In addition, high precision SM predictions also require high precision CKM input. Continued effort to reduce the uncertainties in CKM matrix elements is thus critical, too (see the discussion in the snowmass white paper~\cite{CKM}).    

There is a multitude of rare $b$ and $c$ decay processes that offer a large number of observables which serve as sensitive probes of beyond SM physics. In this white paper we discuss inclusive and exclusive radiative decays of $B$ mesons, purely leptonic decays of $B$ mesons, inclusive and exclusive semileptonic FCNC decays of $B$ mesons and $\Lambda_b$ baryons, rare decays of $b$ hadrons to final states with di-neutrinos, as well as radiative and rare semileptonic decays of charm hadrons. Lepton flavor universality (LFU) tests in rare decays and lepton flavor violation in rare decays are covered in the snowmass white paper~\cite{Guadagnoli:2022oxk}.
The various processes and observables are highly complementary in their sensitivity to new physics. An effective rare decay program thus requires the study of a large number of processes.

Interestingly enough, a number of existing experimental results on rare $b$ decays do not agree well with the corresponding SM predictions. Deviations are observed by LHCb in the angular distribution of the $B \to K^* \mu\mu$ decay, the branching ratios of the $B_s \to \phi \mu\mu$, $B \to K^* \mu\mu$, and $B \to K \mu\mu$ decays, and in the LFU ratios $R_K$ and $R_{K^*}$. If these so-called ``B anomalies'' were to be confirmed to be signs of new physics, it would have a transformative impact on the whole field of particle physics. A completely new mass scale in particle physics could be established and the existence of further new phenomena, in particular the existence of new particles that can be searched for and discovered directly at future particle colliders, would in principle be guaranteed. 

Given the high stakes, we expect that a large part of the rare decay program in the coming years will be dedicated to processes that are related to the $B$ anomalies. On the experimental side, of particular interest will be the continued improvement of results on exclusive rare decays of $B$ mesons by LHCb. An important role will be played by the purely leptonic decay $B_s \to \mu^+\mu^-$ which can be predicted with higher precision than the semileptonic counterparts. Eagerly awaited in the context of the $B$ anomalies are also results from Belle II on the inclusive decay $B \to X_s \ell \ell$. On the theory side, one can expect improved predictions of local form factors from lattice QCD, as well as continued development of methods that allow one to determine long-distance effects in rare decays from data.
Looking beyond the anomalies, the expected experimental results on rare $b \to d$ and $c \to u$ decays will give qualitatively new information on new physics and can provide important insights into the flavor structure of the new physics.  

Whether or not new physics exists in rare decays at an experimentally detectable level, it is highly motivated to push the experimental and theoretical precision beyond the current levels. This will allow us to further extend the indirect reach of rare decays and to explore physics beyond the Standard Model with unprecedented sensitivity. 

\section{Rare decays as probes of new physics} \label{sec:intro}

The exquisite sensitivity of rare decays to new physics has its origin in the peculiar flavor structure of the SM. 
In the SM, the only sources of flavor violation are the hierarchical Yukawa couplings of the Higgs boson to the SM quarks and leptons, implying that FCNC processes of quarks are suppressed by a loop factor and by small CKM matrix elements. Characterizing rare decay processes by an effective interaction analogous to the Fermi constant $G_F$ (the interaction strength of muon decay), order of magnitude estimates of the SM predictions for $b \to s$, $b \to d$, and $c \to u$ decays are indeed tiny
\begin{eqnarray}
 G_{b \to s} &\sim& \frac{\alpha}{4\pi} \frac{m_t^2}{m_W^2} |V_{tb} V_{ts}^*| G_F \simeq 9.6 \times 10^{-5} \times G_F \simeq \frac{1}{(30~\text{TeV})^2} ~, \label{eq:G_bs}  \\
 G_{b \to d} &\sim& \frac{\alpha}{4\pi} \frac{m_t^2}{m_W^2} |V_{tb} V_{td}^*| G_F \simeq 1.9 \times 10^{-5} \times G_F \simeq \frac{1}{(67~\text{TeV})^2} ~, \label{eq:G_bd} \\
 G_{c \to u} &\sim& \frac{\alpha}{4\pi} |V_{cb} V_{ub}^*| G_F \simeq 9.9 \times 10^{-8} \times G_F \simeq \frac{1}{(930~\text{TeV})^2} ~. \label{eq:G_cu} 
\end{eqnarray}
Note that in the case of the $c \to u$ decays, the above short distance estimate is in many cases dominated by long distance (resonance) contributions.
Currently, rare charm decays probe generically scales of a few TeV, which is comparable to the new physics reach of the LHC~\cite{Fuentes-Martin:2020lea}. Existing measurements of rare $b$ hadron decays already probe scales as high as $100$~TeV~\cite{Altmannshofer:2012az, Altmannshofer:2017yso, DiLuzio:2017chi}. 
While the reach of meson mixing observables to heavy new physics is nominally even higher, rare decays offer a multitude of observables that allow new and important tests of the SM and its extensions.
Making optimal use of the high sensitivity of rare decays to new physics requires: (i) experimental measurements of rare decays with as high precision as possible; (ii) theoretical control over the uncertainties in the SM predictions at a comparable level or better. 
If this can be achieved, rare decays can indirectly explore very high mass scales, well beyond the direct reach of collider experiments. 

On the experimental side, the LHC ushered in a new era of exploration of rare decays.
The numbers of $b$ and $c$ hadrons that are produced at $B$ and charm factories are dwarfed by the corresponding numbers at the LHC. 
Over the last decade, many rare decays have been measured with unprecedented precision by LHCb. Many additional decay modes will become accessible in the future runs of the LHC. Complementary information will come from the Belle II experiment that recently started physics data taking.

The challenge on the theory side is to obtain precision predictions for rare decays at the hadronic level. Using effective field theory methods, the description of rare decays can be factorized into a short distance and a long distance part. The short distance physics (the weak interactions in the SM, or the heavy new physics) can be calculated with high precision using perturbation theory. The long distance hadronic physics on the other hand requires reliable non-perturbative methods like lattice QCD.

Overall, the role of the rare decay program is two-fold:
\begin{itemize}
 \item If experimental results agree with SM predictions, strong constraints can be derived on new physics models. In particular, the size of new sources of flavor violation can be strongly restricted which has important implications for new physics model building. A well known example is the $B \to X_s \gamma$ decay. The good agreement between the predicted and observed branching ratio is putting pressure on models with extended Higgs sectors, supersymmetric models, and many other new physics scenarios. Constraints from rare decays and from direct searches at colliders are highly complementary. While rare decays rely on the presence of flavor changing couplings of the new physics, their mass reach is not limited by the available center of mass energy of colliders.
 \item Significant disagreements between the experimental data and SM predictions (i.e. the presence of ``anomalies'') can be interpreted as indirect signs of new physics. Establishing a new physics origin of flavor anomalies would have a transformative impact on the field. The indirect sign of new physics allows one to identify a new mass scale in particle physics which becomes the next target for direct exploration at future high-energy colliders.
\end{itemize}

Since almost a decade, various persistent rare $B$ decay anomalies have created considerable excitement in the community.
For example, the angular distribution of the decay products in the $B \to K^*(\to K \pi) \mu\mu$ decay measured by LHCb shows a $\sim 3 \sigma$ discrepancy with the SM prediction~\cite{LHCb:2020lmf, LHCb:2020gog}. Moreover, the measured branching ratios of the $B_s \to \phi \mu\mu$, the $B \to K^* \mu\mu$, and the $B \to K \mu\mu$ decays are all significantly lower than expected in the SM~\cite{LHCb:2014cxe, LHCb:2016ykl, LHCb:2021zwz}. The mentioned anomalies are supported by hints for lepton flavor universality (LFU) violation~\cite{LHCb:2017avl, LHCb:2019efc, LHCb:2021trn, LHCb:2021lvy}. The SM robustly predicts LFU in rare $B$ decays and the hints for LFU violation are therefore considered theoretically clean signs of new physics. Intriguingly, the hints for LFU violation are fully compatible with the observation of reduced branching ratios and the anomaly in the $B \to K^* \mu\mu$ angular distribution. 
Global fits of rare $B$ decay data find consistently very strong preference for new physics~\cite{Geng:2021nhg, Altmannshofer:2021qrr, Hurth:2021nsi, Alguero:2021anc, Ciuchini:2021smi} and motivated a large model building undertaking, as summarized e.g. in~\cite{Altmannshofer:2022aml}. The new physics scale associated with the $B$ anomalies is possibly within reach of either the LHC or the next generation of colliders~\cite{Allanach:2017bta, Hiller:2018wbv, Huang:2021biu, Asadi:2021gah, Altmannshofer:2022xri, Azatov:2022itm}. If the observed pattern of anomalies should indeed turn out to be a new physics effect, one can expect many discoveries at the high energy and high intensity frontiers in the foreseeable future.

\subsection{The effective Hamiltonian framework}

The language of the effective Hamiltonian is a convenient way to frame the discussion of the rare decay program. 
The physics that leads to rare decays can be model independently encoded in the Wilson coefficients of dimension six operators
\begin{equation}  \label{eq:Heff}
 \mathcal H_\text{eff} = -\frac{4 G_F}{\sqrt{2}} V_{tb}V_{ts}^* \frac{\alpha}{4\pi} \sum_i \Big[ (C_i^\text{SM} + \Delta C_i ) O_i +  \Delta C_i^\prime O_i^\prime \Big] ~,
\end{equation}
with the most important operators
\begin{align}
O_7 &= \frac{1}{e}
(\bar{s} \sigma_{\mu\nu} P_{R} b) F^{\mu\nu}\,,
&
O_7^{\prime} &=  \frac{1}{e}
(\bar{s} \gamma_{\mu} P_{L} b) F^{\mu\nu}\,,
\\
O_9 &=
(\bar{s} \gamma_{\mu} P_{L} b)(\bar{\ell} \gamma^\mu \ell)\,,
&
O_9^{\prime} &=
(\bar{s} \gamma_{\mu} P_{R} b)(\bar{\ell} \gamma^\mu \ell)\,,
\\
O_{10} &=
(\bar{s} \gamma_{\mu} P_{L} b)( \bar{\ell} \gamma^\mu \gamma_5 \ell)\,,
&
O_{10}^{\prime} &=
(\bar{s} \gamma_{\mu} P_{R} b)( \bar{\ell} \gamma^\mu \gamma_5 \ell)\,,
\\
O_{S} &= m_b
(\bar{s} P_{R} b)( \bar{\ell}  \ell)\,,
&
O_{S}^{\prime} &= m_b
(\bar{s}  P_{L} b)( \bar{\ell}  \ell)\,,
\\
O_{P} &= m_b
(\bar{s} P_{R} b)( \bar{\ell} \gamma_5 \ell)\,,
&
O_{P}^{\prime} &= m_b
(\bar{s}  P_{L} b)( \bar{\ell} \gamma_5 \ell)\,.
\end{align}
The leptons $\ell$ can be either electrons, muons, or taus. 
Lepton flavor violating interactions are not included here but are instead discussed in the snowmass white paper~\cite{Guadagnoli:2022oxk}.
The un-primed operators are already present in the SM and we decompose the corresponding Wilson coefficients in a SM part $C_i^\text{SM}$ and a new physics part $\Delta C_i$. The primed operators with Wilson coefficients $\Delta C_i^\prime$ involve right-handed quark currents and describe genuine new physics effects. Complex Wilson coefficients can give rise to CP asymmetries. The expressions above correspond to the Hamiltonian for rare $b \to s$ decays. The Hamiltonians for $b \to d$ and $c \to u$ transitions can be formulated analogously. 
The effective Hamiltonian reflects the aforementioned factorization between the short distance physics (the Wilson coefficients) and the long distance physics (the hadronic matrix elements of the operators).

Theory predictions of rare decay observables can be written as functions of the new physics Wilson coefficients $\Delta C_i^{(\prime)}$. Comparing experimental results to the theory predictions allows one to determine the Wilson coefficients from data. Evidence for a non-zero $\Delta C_i^{(\prime)}$ indicates the presence of new physics. The effective Hamiltonian framework can capture any new physics as long as it is heavy compared to the $b$ or $c$ hadrons. In the presence of new light degrees of freedom (e.g. sterile neutrinos, axions, dark photons, ...) dedicated studies are required.

\subsection{Most promising directions}

To probe each new physics Wilson coefficient with the highest possible sensitivity, a vast array of rare decays needs to be considered. In the following, we will list the most promising processes that are either motivated by the current anomalies or that will give qualitative new information on new physics in rare decays in the coming years. The experimental prospects and the theoretical challenges related to those processes will be discussed in detail in the following sections~\ref{sec:exp} and \ref{sec:theory}. 

Given the high impact that an indirect discovery of new physics in rare decays would have on particle physics, it is of utmost importance to clarify the existing anomalies. At LHCb, lots of focus will therefore be on semi-leptonic electroweak penguin decays $B \to K \mu \mu$, $B \to K^* \mu\mu$, and $B_s \to \phi \mu\mu$. More data and better theory predictions will be required to gain more insight into the anomalies. Additional information will come from \CP asymmetries in these decays modes. In fact, new physics generically contains new sources of \CP violation that could generate observable \CP asymmetries in the rare decays. 

Complementary to the $B$ meson decays are the baryonic decays $\Lambda_b \to \Lambda \mu\mu$. $\Lambda_b$ baryons are ubiquitous at the LHC, and in recent years progress has already been made in the theory description of these decays. Similarly to the meson decays, baryonic decays offer a multitude of observables, including branching ratios, angular distributions, and \CP asymmetries.

Still in the context of the anomalies, of particular interest will be the inclusive decays $B \to X_s \ell\ell$ that can be measured at Belle II. Compared to the exclusive $b \to s \ell\ell$ decays mentioned above, hadronic uncertainties are under better theoretical control. Studies of $B \to X_s \ell\ell$ at Belle II will thus provide important complementary information on the anomalies.

Yet another process that will be crucial to establish a possible new physics origin of the anomalies is the purely leptonic decay $B_s \to \mu\mu$. The $B_s \to  \mu\mu$ decay has a very small branching ratio in the SM of approximately $3\times 10^{-9}$ and it is also theoretically very clean. Thus, it is a highly sensitive probe of new physics~\cite{Altmannshofer:2017wqy}. The observation of $B_s \to  \mu\mu$ at the LHC was an important milestone in the rare $B$ decay program~\cite{CMS:2014xfa, ATLAS:2018cur, CMS:2019bbr, LHCb:2021vsc}. With sufficient data one will also have access to the $B_s \to \mu\mu$ effective lifetime, a complementary probe of new physics in $b \to s\mu\mu$. Proof of principle measurements already exist~\cite{CMS:2019bbr, LHCb:2021vsc}.

Qualitative different information about new physics in rare $B$ decays can be gained from transitions that involve either different quark or lepton flavors, e.g. $b \to d \mu \mu$ or $b \to s \tau \tau$ or $b \to s\nu\nu$ decays. Also $b \to s ee$ decays are obviously relevant in this context, but they are discussed in the snowmass white paper~\cite{Guadagnoli:2022oxk}.
Compared to $b \to s \mu \mu$ decays, the $b \to d \mu \mu$ decays are further suppressed, see equation~\eqref{eq:G_bd}, and rarer by an approximate factor $|V_{ts}/V_{td}|^2 \sim 25$. In the upcoming runs of LHCb, one can expect a substantial amount of $b \to d \mu \mu$ events. Over the coming years the expected statistics will enable precision studies of decays like $B \to \pi \mu\mu$, $B \to \rho \mu\mu$, and $B_s \to K^* \mu\mu$, analogous to the $b \to s \mu\mu$ decays. Also first observation of the extremely rare $B^0 \to  \mu\mu$ decay will become possible.

Rare decays with taus in the final state are very challenging from the experimental perspective. Existing bounds on the branching ratios of decays like $B_s \to \tau \tau$, $B^0 \to \tau \tau$, and $B \to K^{(*)} \tau \tau$ are orders of magnitude above the SM predictions~\cite{BaBar:2005mbx,BaBar:2016wgb,LHCb:2017myy,Belle:2021ndr}. As many models of new physics predict the largest effects for the 3rd generation, further study of $b \to s \tau \tau$ decays is highly motivated. Precision measurements of these decays are required to complete the studies of lepton flavor universality in $b \to s \ell\ell$ and $b \to d \ell \ell$ decays.

Beyond the leptonic and semi-leptonic decays discussed so far, radiative decays are another class or rare $B$ decays that will remain highly relevant in the future. The radiative $b \to s \gamma$ modes are particularly sensitive probes of the dipole operators $O_7$ and $O_7^\prime$. Complementary information on the corresponding Wilson coefficients $C_7$ and $C_7^\prime$ can be gained from a variety of observables, including the branching ratio of the inclusive $B \to X_s \gamma$ decay, the direct CP asymmetry $A_\text{CP}$ and the time dependent CP asymmetry $S_{K^*\gamma}$ in the exclusive $B \to K^* \gamma$ decay, as well as the angular distribution of the $B \to K^* ee$ decay at low $q^2$. It would be worthwhile to extend those studies to the corresponding $b \to d \gamma$ decays as well.

An additional interesting branch of studies is the the one focused on rare FCNC charm decays. These decays represent an unique probe for the presence of new physics in the up-quark sector and a relatively unexplored area of research. Due to the GIM and CKM suppression, short distance contribution in $\vert \Delta \cquark \vert = \vert \Delta \uquark \vert = 1$ transitions are more effectively suppressed compared to the down-quark counterparts. Moreover, CP asymmetries are also CKM-suppressed and the very effective GIM mechanism and consequently the absence of the axial-vector lepton current leads to a very specific angular distributions on the final states particles. Despite the lack of a robust effective theory framework to deal with the non-perturbative dynamics of rare charm decays, the SM symmetries lead to a very unique phenomenology and allow to define clean null-test observables.

{\def\arraystretch{1.5}
\setlength\tabcolsep{12pt}
\begin{table}[tb] \centering
\begin{tabular}{lcccc}
 \hline\hline
 & $C_7, C_7^\prime$ & $C_9, C_9^\prime$ & $C_{10}, C_{10}^\prime$ & $C_{S,P}, C_{S,P}^\prime$ \\ \hline
 $B \to X_s \gamma$ & $\bigstar$ & & &  \\  \hline
 $B \to K^* \gamma$ & $\bigstar$ & & &  \\  \hline
 $B_s \to \phi \gamma$ & $\bigstar$ & & &  \\  \hline
 $B \to K^* ee$ at very low $q^2$ & $\bigstar$ & & &  \\  \hline
 $B \to X_s \mu\mu$ & $\bigstar$ & $\bigstar$ & $\bigstar$ & $\star$ \\  \hline
 $B \to K^{(*)}\mu\mu$ & $\bigstar$ & $\bigstar$ & $\bigstar$ & $\star$ \\  \hline
 $B_s \to \phi \mu\mu$ & $\bigstar$ & $\bigstar$ & $\bigstar$ & $\star$ \\  \hline
 $\Lambda_b \to \Lambda \mu\mu$ & $\bigstar$ & $\bigstar$ & $\bigstar$ & $\star$ \\ \hline
 $B_s \to \mu\mu$ & & & $\bigstar$ & $\bigstar$ \\
 \hline\hline
\end{tabular}
\caption{Complementary sensitivity of various $b \to s$ decays to the Wilson coefficients of the effective Hamiltonian. (The small star indicates that there is in principle sensitivity but other processes are much more sensitive.)}
\label{tab:WCs}
\end{table}
}

While this is a long list of relevant processes and observables, we stress that all those processes are highly complementary in their sensitivity to new physics. In the context of the effective Hamiltonian in equation \eqref{eq:Heff}, each of the 
mentioned processes depends in a characteristic way on a specific subset of Wilson coefficients. 
This is illustrated in table~\ref{tab:WCs} for the case of rare decays based on the $b \to s$ transition. The combination of all processes is required to obtain a complete picture of new physics in rare decays.

\section{Experimental opportunities} \label{sec:exp}

\subsection{Current and future machines}

In this section the current and future machines devoted to the study of rare $b$ and $c$ hadrons rare decays are presented.

\subsubsection{Belle II}

The \belleii $4\pi$ detector is located at the upgraded \superkekb \epem collider at the KEK laboratory~\cite{Forti:2022mti}. 
The two asymmetric beams have a total energy in the center of mass equal to the mass of the $\Upsilon(4S)$ resonance and the $B\overline{B}$ pair produced in its decay are boosted in the forward direction. \belleii started operating in 2021 with a partial vertex detector foreseen to be installed in 2023~\cite{BelleII:2022snowmass} and targeting to collect a sample of 50\invab of \epem collisions, corresponding to $\sim 50\times 10^9$ $B\overline{B}$ pairs produced. The lower cross section for the production for $B\overline{B}$ mesons, $\sigma_{B\overline{B}}\sim 1\nb$ is compensated by a very clean environment where the initial state is know allowing the measurement of rare decays with neutrinos in the final state or possibly new undetectable particles. Moreover, working at the $\Upsilon(4S)$ resonance, fewer initial states are accessible compared to experiments at hadronic machines. 

\subsubsection{\lhcb}

The \lhcb experiment is a single-arm forward spectrometer at the Large Hadron Collider designed to study the properties of particles containing \bquark and \cquark~\cite{LHCb:2008vvz}. Differently from lepton colliders, hadron colliders create a large number of elementary particles for each collision. Even if $b\overline{b}$ pairs are produced with a large cross section ($\sigma_{b\overline{b}}\sim 300-600\mub$) the inelastic scattering has a cross section 200 times larger of $\sigma_{b\overline{b}}$ and the initial state is unknown.
In this harsh environment, $b$ hadrons are produced with a very large boost allowing to reduce the large background contribution coming form the primary vertex.
Given the large $c\overline{c}$ cross section, the \lhcb experiment can be also considered a charm factory allowing to probe up-type FCNC $\decay{c}{u\ell^+\ell^-}$ processes.

Differently from electron-positron collider, all species of \quark hadrons are produced at hadronic machines, making it possible to access \Bs mesons and baryonic rare decays.

During the first two Runs of LHC (Run1 and Run2), the LHCb experiment collected $9\invfb$ of $pp$ collisions and it was able to uncover several interesting measurements on flavour physics in the area of rare decays, commonly referred to as flavour anomalies.

The LHCb experiment will restart operation in 2022 with an upgraded detector and trigger (Upgrade Ia) and is expected to to collect an additional 50\invfb over approximately 4 years. Further upgrades called Upgrade Ib and Upgrade II are planned with the hope of collecting a total 300\invfb data set~\cite{LHCb:2018roe}.  

\subsubsection{\atlas and \cms}

\atlas and \cms are the two large general purpose experiments of the LHC~\cite{ATLAS:2008xda, CMS:2008xjf}. These two detectors are designed to operate at a larger luminosity compared to the LHCb experiments covering a different rapidity region. These detectors have dedicated $B$ triggers relying on high-$p_T$ selection on leptons and no dedicated hadronic particle identification capabilities. For this reason their {\it flavour} program is limited compared to $B$-physics experiments and mostly focused to decays with muons in the final state. 
At the moment, the two experiments have collected only 5\% of the total integrated luminosity envisaged to be delivered in the LHC lifetime. The high luminosity upgrade of the collider (HL-LHC) will allow to increase the current data-set by a factor $\sim 20$ reaching an expected integrated luminosity of 3000\invfb of $pp$ collisions~\cite{ATL-PHYS-PUB-2022-018}.

\subsubsection{\besiii}

The \besiii spectrometer operates at the Beijing Electron Positron Collider (BEPCII), and uses \epem collision with a center of mass energy ranging from 2.0 and 4.7\tev~\cite{BESIII:2009fln}. Given the energy spectrum accessible, its physics program is mainly focused on charm and tau physics. Since the start of the operation in 2009, \besiii has collected $\sim 30\invfb$ of data. Its physics program is foreseen to continue for at least the next 5-10 years~\cite{BESIIISnowmass}.

\subsubsection{Future machines}

$Z$-factories, as the one proposed for FCC-$ee$~\cite{Bernardi:2022hny} or CEPC~\cite{Cheng:2022zyy}, are a great opportunity for flavour physics studies. Even if the expected number of \bquark-hadrons are relatively small compared with the experiments at the hadronic machines, $Z$-factories can rely on a negligible pile up, a good geometric coverage, a good reconstruction efficiency and a good resolution on missing momentum thanks to the inclusive reconstruction of the decay tree.
The latter characteristic allows to access decays such as $\bquark \to \squark \nu \overline{\nu}$ processes, where the outgoing neutrinos only manifest themselves as missing momenta. Compared to the $B$ factories such as \belleii, the \bquark-hadrons receive an higher boost at the $Z$ pole leading to a more accurate tracking reconstruction and a lower contamination from heavy flavour decays.

\subsubsection{Comparison between current and future machines}

In general, hadronic machine experiments offer the opportunity to access to a broader set of heavy hadrons, on the other hand a they have to cope with a larger background contamination. Experiments at \epem colliders work with a cleaner environment at lower occupancy and are able to fully reconstruct the decay kinematics. As shown in Tab.~\ref{tab:bproduction}, it is clear that these advantages are well compensated by the hadronic machines once the $B$ meson production rate is considered.

\begin{table}[tbh]
    \centering
    \begin{tabular}{c|c|c|c}
         Channel & \text{\belleii} & \text{LHCb-U1a} & $Z$-factory \\
         \hline
         \Bz, \Bzb& $\sim 5\times 10^{10}$ & $\sim 6\times 10^{13}$ & $\sim 1.2\times 10^{11}$\\
         \Bpm & $\sim 5\times 10^{10}$ & $\sim 6\times 10^{13}$ & $\sim 1.2\times 10^{11}$\\
         \Bs, \Bsb & $\sim 6\times 10^{8}$ & $\sim 2\times 10^{13}$ & $\sim 3.2\times 10^{10}$\\
         \Bcpm & $-$ & $\sim 2\times 10^{11}$ & $\sim 2.2\times 10^{8}$\\
         \Lb, \Lbbar & $-$ & $\sim 2\times 10^{13}$ & $\sim 1.0\times 10^{10}$\\
    \end{tabular}
    \caption{Number of \bquark-hadrons expected to be produced at \belleii, \lhcb and future $Z$-factory as FCC-$ee$. \belleii expected yields are evaluated considering that the experiment will run at \FourS and \FiveS with an integrated luminosity of 50\invab and  5\invab respectively. \lhcb expected yields are evaluated at 50\invfb considering \bquark-hadrons to be produced in the detector acceptance. $Z$-factory expected yields are taken from Ref.~\cite{Snowmass:cepc}}
    \label{tab:bproduction}
\end{table}

\begin{sidewaysfigure}
    \centering
    \includegraphics[width=\textwidth]{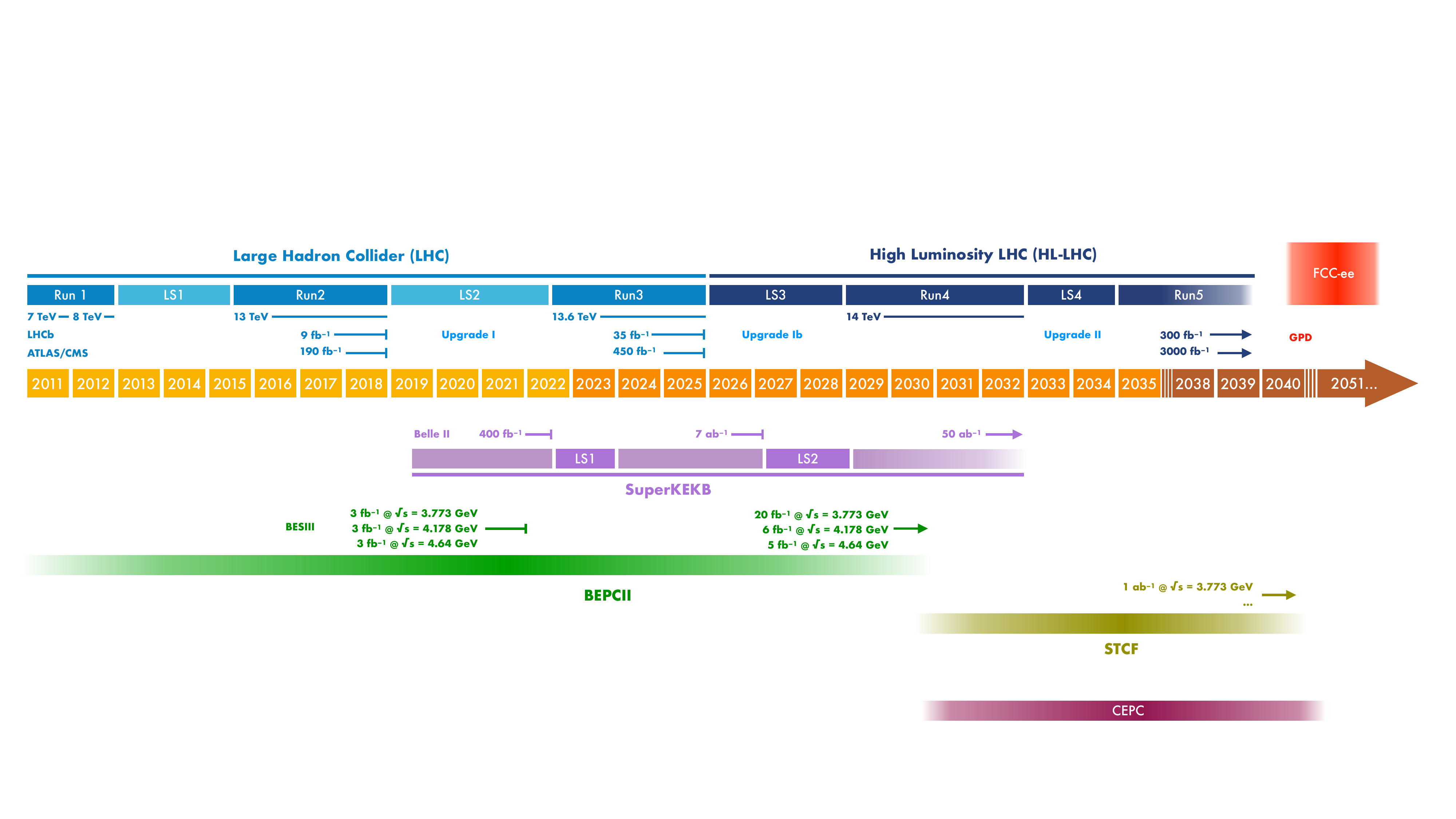} \\[16pt]
    \caption{Timelines of the main experiments performing precision measurements on rare \bquark and \cquark processes. The integrated luminosities already collected and expected are taken from Refs.~\cite{LHC:timeline,FCC:timeline,BelleII:newtimeline}. FCC-$ee$ is placed in the same row of the LHC timeline since this project can limit the lifetime of the LHC datataking. CEPC collider expected timeline is taken from Ref.~\cite{CEPC:timeline}. \besiii experiment timeline and future tau-charm factory timelines relevant for the charm physics program are taken from Ref.~\cite{BESIII:2020nme} and Ref.~\cite{STCF:timeline} respectively.}
    \label{fig:timeline}
\end{sidewaysfigure}

The timeline shown in Fig.~\ref{fig:timeline}, gives an idea of the role played by the different experiments contributing to the search for new physics with FCNC processes of \bquark and \cquark flavored hadrons in the short and long term period.

\subsection{Current experimental status and expected sensitivities}

In this section the current experimental status of the study of $B$ and $D$ rare decays and expected future sensitivities are reported.

\subsubsection{Purely leptonic $B$ decays}

As reported in Sec.~\ref{sec:th:leptonic}, purely leptonic rare $B^0$ and $B^0_s$ decays are very well predicted in the SM. While the \bsmumu decays have been already observed at the LHC, the other two-body decays, due to the additional helicity or CKM suppressions, as for the \Bsee decays or the \Bdmumu decays, or the technical challenge they represent at hadronic machines, such as for the \Bstautau decays, only upper limits on their branching fractions have been set. 

The branching fraction of the \mbox{\Bsmm} decay measured by the \lhcb collaboration is \mbox{$\BRof \Bsmumu=\Bsbr$}, where the systematic uncertainty is dominated by the uncertainty on the ratio of fragmentation fractions $f_s/f_d$ ($\sim 3\%$) and the uncertainty on \BuJpsiK branching fraction used as normalisation channel. This represents the most precise single-experiment measurement and it is based on the 9\invfb collected during the first two runs of LHC.
Within the same analysis the branching fraction of \bdmumu decays was also measured to be \mbox{$\BRof \Bdmumu=\Bdbr$} with a statistical significance 1.7$\sigma$. Since no evidence of \bdmumu decays is reported a upper limit on its branching fraction is evaluated to be $\BRof \Bdmumu < \mbox{$\Bdobslimitnf$}$ at 95\% CL. 

The ratio of \bsmumu and \bdmumu branching fractions, which is a powerful observable to test minimal flavour
violation, has been also measured by the \lhcb experiment and found to be \RmumuLHCb. Even if this quantity is statistically limited it will be probed at the 10\% level with 300\invfb~\cite{LHCb:2018roe}.

Also the \atlas and \cms experiments have measured the \bdsmumu processes branching fractions with a fraction of the data collected at LHC during the first two runs.
The latest \atlas collaboration results are based on the data samples corresponding to
integrated luminosities of 25~\invfb collected in 2011 and 2012, and 26.3~\invfb collected in 2015 and 2016. The ATLAS analysis yields $\BRof\Bsmumu = \ATLASBsmumuMeasure$ and $\BRof\Bdmumu = \ATLASBdmumuMeasure$ with a significance for the \Bsmumu signal of \ATLASBsmumuSignificance $\sigma$. A 95\,\% confidence level (CL) upper limit for the \Bdmumu signal is $\BRof\Bdmumu < \ATLASBdmumuUpperLimit$. The main contribution to the total uncertainty comes from the statistical uncertainty. 

The \cms results are based on a data sample corresponding to integrated luminosities of 25 \invfb collected in 2011 and 2012, and 36 \invfb collected in 2016. The measured branching fractions are~\cite{CMS:2019bbr} $\BRof\Bsmumu = \CMSBsmumuMeasure$ and
$\BRof\Bdmumu = \CMSBdmumuMeasure$
with a signal significance of $\CMSBsmumuSignificance\,\sigma$ and $\CMSBdmumuSignificance\,\sigma$, respectively. 
In the quoted \Bsmumu branching fraction measurement, the first uncertainty combines the 
experimental statistical and systematic uncertainties on the measurement, while the second is due to the uncertainty in the ratio of fragmentation fractions $f_d/f_s$. A 95\,\% CL upper limit for the \Bdmumu decay is evaluated to be $\BRof\Bdmumu < \CMSBdmumuUpperLimit$.

A combined result using the measurements from \atlas \cms and \lhcb based on Run1 data and a fraction of Run2 data set has been also obtained~\cite{LHCb:internalnote} with the same precision of the latest \lhcb result with the full Run1 and Run2 data sets. Assuming the current analysis performances, a new combination of the analyses of the three LHC experiments with the full Run1 and Run2 data samples will allow to reach a statistical uncertainty on the \bsmumu branching fraction of $\sim 7\%$ which can be reached by \lhcb experiment only at the end of Run4.

A summary of the current status and the extrapolated sensitivities of the LHC experiments are reported in Tab.~\ref{tab:b2mumusens}

\begin{table}[tb]\centering
\caption{Summary of the current and expected experimental precision for \bsmumu and \bdmumu observables. The expected uncertainty are reported for \lhcb at 23\invfb (LHCb-U1a) and 300\invfb (LHCb-U2) while for ATLAS and CMS are evaluated at 3\invab.}\label{tab:b2mumusens}
\begin{tabular}{l|ccccc}
\text{Observable} & \text{Current} & \text{LHCb-U1a} & \text{LHCb-U2} & \text{ATLAS} & \text{CMS}\\
\hline
\BRof\bsmumu  $(\times 10^9)$& \BsmumuCurrent & \BsmumuLHCbUI & \BsmumuLHCbUII & \BsmumuATLASHL & \BsmumuCMSHL \\
$\frac{\BRof\bdmumu}{\BRof\bsmumu}$ & $\sim\RmumuCurrent$ & $\sim\RmumuLHCbUI$ & $\sim\RmumuLHCbUII$ & $\RmumuATLASHL$ & $\sim\RmumuCMSHL$\\
$\tau_{\mu\mu}$ & $\sim\tauCurrent$ & \taumumuLHCbUI & \taumumuLHCbUII & \taumumuATLASHL & \taumumuCMSHL \\
\end{tabular}
\end{table}

An additional observable, already accessible by CMS and LHCb experiments, is the effective lifetime of the \Bsmumu, $\tau_{\mu\mu}$, which is simply defined as the mean lifetime of \Bsmumu decays. 
This quantity would allow to break the degeneracy between any possible contribution from new scalar and pseudoscalar mediators.
In its recent analysis~\cite{LHCb:2021vsc,LHCb:2021awg}, the LHCb collaboration measured $\tau_{\mu\mu} = 2.07 \pm 0.29$ which is consistent with the expectation~\cite{PDG2020} at 2.2$\sigma$ level and the most precise measurement today. The relative uncertainty of this quantity is expected to decrease to approximately to 8\% with 23\invfb and 2\% with 300\invfb of data. 

With the additional data that will be available during the future LHC runs, the LHCb experiment will be 
able to also measure the time-dependent CP asymmetry of \Bsmumu decays. Assuming a tagging power of about 3.7\%, a dataset of 300\invfb allows a pure sample of more than 100 flavour-tagged \Bsmumu decays to be reconstructed and their time-dependent CP asymmetry to be measured and the related CP-violating term $\mathcal{S}_{\mu\mu}$ with an uncertainty of about 0.2. 
On the other hand, the signal yield expected in the end of LHC Run3 23\invfb data set is too low to allow a meaningful constraint to be set on $\mathcal{S}_{\mu\mu}$. Any nonzero value for $\mathcal{S}_{\mu\mu}$ would automatically indicate evidence of CP-violating phases beyond the SM.

It can be safely stated that the three experiments at LHC will dominate the field in the measurement of \Bsmumu decay properties for long time. 

Using the full Run1 and part of Run2 data sets, the \lhcb collaboration set the most stringent limits on \Bsee and \Bdee branching fractions, $\BRof\Bsee < \Bseelimit$ and $\BRof\Bdee < \Bdeelimit$ at 95\% CL~\cite{LHCb:2020pcv}, which are five orders of magnitude above the SM expectations. Even if the SM rates are unreachable with the full statistics that \lhcb will have collected after the Upgrade II, the search for \Bsee decays can already probe possible NP scenarios where the helicity suppression is lifted by new (pseudo-)scalar contributions~\cite{Fleischer:2017ltw}.


The \lhcb collaboration also performed a search for \Bstautau and \Bdtautau decays using the first 3\invfb of data collected during Run1. Using the three-prong decays of $\tau$ leptons, LHCb set the upper limits~\cite{LHCb:2017myy} of $\BRof\Bstautau < \Bstautaulimit$ at 95\% CL and $\BRof\Bdtautau < \Bdtautaulimit$ at 95\% CL. Even if these represent the most stringent limits today, they are about five orders of magnitude above the SM expectations. Based on the current anomalies, an enhancement of the rate of these decays close to the current experimental reach makes the search of these decays still interesting, even though, a measurement of the SM value with the current experimental precision will remain out of the reach of LHCb. Naively scaling the limits with the integrated luminosity, the limit on \BRof\Bstautau is expected to improve by a factor $\sim 14$ by the end of Upgrade II. Also, \belleii is expected to have a competitive limit on \Bdtautau decays setting a limit on the branching fraction to be $< 9.6 \times 10^{-5}$ with 50\invab~\cite{Belle-II:2018jsg}.

\subsubsection{Semileptonic decays}

Compared with the purely leptonic $B$ decays, semileptonic decays have relatively higher branching fractions and offer a large set of observables, such as angular observables or differential branching fractions expressed as a function of the dilepton invariant mass squared ($q^2$). 

The \lhcb experiment found that the measurement of several exclusive \bsll decay branching fractions were systematically lower than their corresponding SM predictions. The largest tension appeared for \BRof\BsToPhimm which is of $\sim 3.6\sigma$ in the $1 < q^2 < 6 \gevgevcccc$ region~\cite{LHCb:2021zwz} with respect to a precise SM prediction. With the future analyses performed by the \lhcb collaboration, the precision of the measurement of the branching fractions will be limited by the knowledge of the $\decay{\B}{\jpsi X}$ decay modes that are used to normalise the observed signals. The \belleii collaboration will certainly play an important role to improve the knowledge of these branching fractions. 

The \belleii collaboration is expected to provide and independent check of the anomalies observed by the \lhcb collaboration using \BdToKstmm and \BdToKstee decays and be competitive with samples of 5\invab to 10\invab of data. 

In any case, the comparison between the predicted and measured branching fractions will be limited by the theoretical knowledge of the local and non-local form factors, even if in the future these are determined parametrically using data driven methods. 

A better way to reduce the theoretical and the experimental uncertainty is to compare regions of angular phase-space of \bsll decays. The angular distribution can be expressed in terms of $q^2$-dependent angular coefficients that depend on the Wilson coefficients and the form-factors. The \lhcb measurements of angular observables in \BdToKstmm decays shown a discrepancy with respect to the SM predictions and, in particular, this is largest in the observable $P_5^\prime$~\cite{LHCb:2020lmf} which has reduced dependence on form factors. With the large data set that the \lhcb experiment will collect with Upgrade II, corresponding to $\sim 440000$ fully reconstructed \BdToKstmm decays, it will be possible to make a precise determination of the angular observables in narrow bins of $q^2$ or using unbinned approaches~\cite{LHCb:2018roe}.

Using the Run 1 and Run 2 data sets, \lhcb was able to observe the decays \BuTopimm and \LbToPpimm. The Upgrade II data set will give the possibility to make precise measurements of many \bdll transitions. Thanks to the mass resolution of the \lhcb detector, the \BsToKstmm decay can be separated from \BdToKstmm and a full angular analysis will be possible~\cite{LHCb:Snowmass}.

Also the \cms and \atlas experiments are expected to contribute to the understanding of the angular quantities of \BdToKstmm decays. With the statistics collected at HL-LHC, the \cms experiment is expected to fully reconstruct $\sim 700000$ \BdToKstmm events, and improve the uncertainty on the $P_5^\prime$ observable by a factor of 15 compared to its previous measurement performed using only 20\invfb at 8\tev. On the other hand, the \atlas experiment is expected to have an improvement that goes from a factor 5 to a factor 9 depending on the muon $p_T$ thresholds used in the different trigger scenarios assumed~\cite{ATL-PHYS-PUB-2022-018}. 

Heavy \bquark baryons are produced copiously at LHC and their rare decays are a prerogative of the \lhcb experiment. The study of $\decay{\Lb}{\Lambda(\to \proton \pip) \mup\mun}$ decays offers an experimental advantage compared with the $B$ meson decays, since the presence of a long-lived baryon in the final state makes these transitions less affected by partially reconstructed background. 
The differential branching fraction of $\decay{\Lb}{\Lambda \mup\mun}$ was measured by \lhcb using just the Run 1 data set~\cite{LHCb:2015tgy}. The main limits of this measurement were the statistics of the data sample and the uncertainty on the normalisation channels used. This will certainly be improved by the \lhcb collaboration with the future analyses. 
Moreover, an analysis of the angular distribution was performed using data collected with the \lhcb detector during the first two runs of LHC showing no deviation from the SM~\cite{LHCb:2018jna}. No projections are available for the future analyses.

The study of decays with excited $\Lambda$ baryons in the final states are experimentally challenging due to the rich resonant structure of the $\Lambda$ decay products. Also in this case \lhcb demonstrated to be the main player with the determination of the CP asymmetries using $\decay{\Lb}{\proton \Km \mup\mum}$ decays~\cite{LHCb:2017slr} and the LFU test with $\decay{\Lb}{\proton \Km \ell^+\ell^-}$ decays~\cite{LHCb:2019efc}. Projection of the latter can be found in the snowmass white paper~\cite{Guadagnoli:2022oxk}.  


Complementary to the measurement of exclusive \bsll decay properties, the measurement of inclusive $B \to X_s \ell^+\ell^-$ decay observables represents an important tool to cross-check the anomalies found by \lhcb. The reconstruction of inclusive decays is usually performed through the sum of exclusive mode method in which the hadronic system, $X_s$, is reconstructed from the pions and kaons in the final states. Therefore, these analyses are quite challenging for experiments like \lhcb, due to the detector geometry and the production mechanism that does not allow to fully reconstruct the decay. This is not an issue at the \epem collider, thanks to the possibility to tag the other $B$ involved in the decay.
The most precise measurement of the partial branching fraction today are from a combination of \belle and \babar measurements, with a relative uncertainty of around 23-21\% depending on the $q^2$ region. To be noted that the theoretical uncertainty in the region [1-6]\gevgevcccc of $q^2$ is 4 times smaller than the measured value. \belleii experiment will be able to further improve this measurement with finer $q^2$ binning, reaching an uncertainty of 6.6-4.7\% depending on the $q^2$ region using a sample of 50\invab~\cite{Belle-II:2018jsg}.   

The forward-backward asymmetry $A_{\rm FB}$ is expected to give stringent constraints on NP contribution to Wilson coefficients $C_9$ and $C_{10}$. The expected uncertainty on $A_{\rm FB}$ for \belleii experiment with 50\invab is 3.1-2.4\% depending on the $q^2$ region considered~\cite{Belle-II:2018jsg}. \belleii experiment is also expected to probe for new \CP violation sources through the measurement of the \CP asymmetries in $B \to X_s \ell^+\ell^-$ decays with a precision of the order of few percents~\cite{Belle-II:2018jsg}. 

\subsubsection{Di-neutrino decays}

Searches for the di-neutrino decays $B \to K \nu\bar\nu$ and $B \to K^* \nu\bar\nu$ have been performed at the $B$ factories using both hadronic and semi-leptonic tagging.
The obtained constraints on the branching ratios from BaBar and Belle~\cite{BaBar:2013npw,Belle:2013tnz,Belle:2017oht} are only a factor of few above the respective SM predictions.

Belle II is expected to make first observation
of the di-neutrino decays. With a data set of $50$\,ab$^{-1}$ and assuming SM-like branching ratios, Belle II can perform a measurement of the $B \to K \nu\bar\nu$ and $B\to K^* \nu\bar\nu$ branching ratios with an uncertainty of approximately $10\%$~\cite{Belle-II:2018jsg}. Also a measurement of the $K^*$ longitudinal polarization fraction $F_L$ with an uncertainty of $\sim 0.11$ will be possible~\cite{Belle-II:2018jsg}.
A first bound on the $B^+ \to K^+ \nu\bar\nu$ branching ratio from Belle II with $63$\invfb, BR$(B^+ \to K^+ \nu\bar\nu) < 4.1 \times 10^{-5}$~\cite{Belle-II:2021rof}, is already competitive with the previous results from \babar and \belle with only a tenth of the previous B-factory data set. 

Given the expected number of $B$ mesons produced at $Z$-pole runs of future circular $e^+ e^-$ colliders like FCC-ee or CEPC, one can expect further improvements of these measurements. This would be welcome as the decays are under good theoretical control (see discussion in section~\ref{sec:th:dineutrino}). $Z$ pole machines also have the unique opportunity to observe the related decay modes of heavier $b$ hadrons $B_s \to \phi\nu\bar\nu$, $\Lambda_b \to \Lambda\nu\bar\nu$, and even $B_c \to D_s \nu\bar\nu$. For example, a measurement of the $B_s \to \phi \nu\bar\nu$  branching ratio with a precision of $\mathcal{O}(1\%)$ might be possible at FCC-ee or CEPC~\cite{Bernardi:2022hny, Cheng:2022zyy, Li:2022tov}.

\subsubsection{Radiative decays}

\belleii can access a wide range of \bsgamma and \bdgamma modes and \CP observables in \Bd decays. Moreover, it has the unique capability to study these transitions both inclusively and using specific exclusive channels. 
The precise and reliable SM prediction of the inclusive $\decay{B}{X_\squark \gamma}$ rate makes this channels sensitive probe for NP. The main source of uncertainty for these measurements at \belleii arises from neutral hadrons faking photons. Depending on the possible scenarios the precision of this measurement will become limited by the systematic uncertainty at approximately 10\invab or at the end of \belleii data taking~\cite{BelleII:2022snowmass}.

Exclusive radiative decays are experimentally more straightforward than the inclusive processes and can be accessed also at hadronic machines. However, absolute rates suffer from larger theoretical uncertainties. These transitions are challenging at \lhcb due to the mass resolution and the contamination from \piz decays. 
However, the \lhcb collaboration played an important role in this sector performing some remarkable measurement such as the determination of the \CP-violating and mixing-induced observables in \BsToPhigamma~\cite{LHCb:2019vks}, the measurement of \CP asymmetry for \BdToKstgamma~\cite{LHCb:2012quo} decays and the first direct observation of the photon polarization in the \bsgamma transition using \BuToKpipigamma decays~\cite{LHCb:2014vnw}. 
The Upgrade II detector is expected to improve the mass resolution for the \piz thanks to a better spatial segmentation, which would be crucial to march with the expected statistical precision of several of these measurements. 
\belleii is expected to have the highest sensitivity in the next decade reaching percent or sub-percent precision on related observables.~\cite{BelleII:2022snowmass}

Weak radiative decays of \bquark baryons are largely unexplored. They offer a unique sensitivity to the photon
polarisation through the study of their angular distributions. \lhcb already demonstrated to be able to access these channel with the measurement of the photon polarisation in \LbToLgamma decays using only a part of data collected during Run II~\cite{LHCb:2019wwi}. Such decays will constitute one of the main topics in the radiative decays programme in the \lhcb Upgrade II.

The angular analysis of \BdToKstee at very low $q^2$ can also give access to the structure of the $b \to s \gamma$ interaction via semileptonic \bsll transitions. 
The recent \lhcb result~\cite{LHCb:2020dof} is consistent with SM predictions and can be used to measure
both the real and imaginary parts of the \BdToKstgamma photon polarisation with a precision
of 5\%. This result makes it possible to constrain the \bsgamma photon polarisation with significantly better precision than the combination of previous measurements.
The experimental conditions at Upgrade II should improve for the \lhcb experiment thanks to a reduced material budget near the $B$ decay vertex, and a precision at the level of 2\% is expected~\cite{LHCb:2018roe}. 

The large data set available at Upgrade II will allow to have also access also to \bdgamma processes where a larger \CP asymmetry is expected. Moreover, \bdee decays at low-$q^2$ will represent an alternative way to measure photon polarisation. 

\subsubsection{Rare charm decays}
The large production rate of charmed hadrons at the LHC has allowed the \lhcb experiment to perform a broad set of fundamental measurements in the charm system. 

The physics reach of this program will be significantly enhanced with the much larger data sets expected from the operation of the Upgrade I and Upgrade II detectors. The fully software trigger of the upgraded experiment
will allow for the collection of orders of magnitude larger samples of charmed hadrons than any other experiment, including \belleii. 

We expect precision measurements of \CP asymmetries, angular analyses and tests for lepton
universality in semi-leptonic decays of charmed hadrons at the percent level~\cite{LHCb:2018roe, BelleII:2022snowmass}, and even below in resonance-dominated regions of di-lepton mass for some semi-leptonic final states. 

Aslo the \besiii experiment will play a significant role in the investigation of rare charm decays with electrons in the final state~\cite{BESIIISnowmass}. 

The \belleii collaboration will be capable to investigate angular distributions of rare radiative
decays but experimental results are not yet available. Efforts including channels with photons and electrons
are also expected to be intensified by LHCb in the future.

\section{Theory challenges} \label {sec:theory}

\subsection{Leptonic $B_s \to \ell \ell$ and $B^0 \to \ell \ell$ decays} \label{sec:th:leptonic}

The purely leptonic decays $B_s \to \ell \ell$ and $B^0 \to \ell \ell$ are the simplest exclusive FCNC decays of $B$ mesons and theoretically very well understood. Up to small QED corrections~\cite{Beneke:2019slt}, non-perturbative QCD effects are parameterized by the $B$ meson decay constants, which are known with sub percent precision from lattice QCD, $f_{B_s} = 230.3(1.3)$\,MeV and $f_{B^0} = 190.0(1.3)$\,MeV~\cite{Aoki:2021kgd, Dowdall:2013tga, ETM:2016nbo, Bazavov:2017lyh, Hughes:2017spc, Boyle:2022uba}.

The branching ratios are helicity suppressed and therefore very small. The SM predictions for the muonic decays are~\cite{Buchalla:1993bv, Bobeth:2013uxa, Beneke:2019slt} 
\begin{equation} \label{eq:Bsmumu_SM}
 \text{BR}(B_s \to \mu \mu)_\text{SM} =  (3.66 \pm 0.14)\times 10^{-9} ~,\quad \text{BR}(B^0 \to \mu \mu)_\text{SM} =  (1.03 \pm 0.05)\times 10^{-10} ~.
\end{equation}
In both decays, the by far dominant uncertainty in the SM predictions is coming from CKM matrix input, in particularly $V_{cb}$. 

Going beyond the SM, these decays are known to be highly sensitive probes of new physics~\cite{Altmannshofer:2017wqy}. In particular, the $B_s \to \mu \mu$ branching ratio provides a theoretically clean probe of the Wilson coefficients $C_{10}^{(\prime)}$. Moreover, the leptonic decays are uniquely sensitive to the scalar and pseudoscalar Wilson coefficients $C_S^{(\prime)}$ and $C_P^{(\prime)}$ as these coefficients lift the helicity suppression. Very interesting are also the decays into taus. While measuring the branching ratio of $B_s \to \tau \tau$ is very challenging experimentally, many BSM scenarios that are motivated by the current $B$ anomalies predict characteristic enhancements of this decay.

Note that the predictions for the $B_s \to \mu \mu$ and $B^0 \to \mu \mu$ branching ratios in equation \eqref{eq:Bsmumu_SM} use the $V_{cb}$ value from inclusive determinations $|V_{cb}|_\text{incl.} = (42.00\pm 0.64)\times 10^{-3}$ from~\cite{Gambino:2016jkc}. Using instead the exclusive value, e.g. $|V_{cb}|_\text{excl.} = (39.4\pm 0.8)\times 10^{-3}$ from~\cite{PDG} would give branching ratios that are more than 10\% smaller (see e.g.~\cite{Buras:2022wpw}). 
While improving the lattice predictions for the decay constants would be clearly welcome, resolving the tension between the inclusive and exclusive $|V_{cb}|$ determinations~\cite{PDG, Ricciardi:2021shl, Altmannshofer:2021uub} will be most important to improve the uncertainties of the theory predictions and to fully make use of the exquisite new physics sensitivity of $B_s \to \ell \ell$ and $B^0 \to \ell \ell$.
Alternatively, one could consider suitable ratios with other observables~\cite{Buras:2003td, Bobeth:2021cxm, Buras:2021nns}, e.g. the $B$ meson oscillation frequencies $\Delta M_d$ and $\Delta M_s$ to remove the sensitivity to the CKM input. Such ratios are sensitive probes of new physics but they introduce dependence on additional hadronic parameters and potentially on additional unrelated new physics.

There are also additional observables in the leptonic decays that are sensitive to new physics. Thanks to the sizable width difference of the neutral $B_s$ mesons, the effective lifetime in $B_s \to \mu \mu$, $\tau_\text{eff}$~\cite{DeBruyn:2012wk}, provides an interesting complementary probe of scalar and pseudoscalar new physics. Existing results on $\tau_\text{eff}$ already start to give non-trivial constraints~\cite{Altmannshofer:2021qrr}. The time dependent CP-asymmetry $S_{\mu\mu}$~\cite{Buras:2013uqa} could give access to imaginary parts of the scalar and pseudoscalar Wilson coefficients.
Also the decay with an additional photon, $B_s \to \mu\mu \gamma$ provides interesting complementary new physics sensitivity~\cite{Guadagnoli:2017quo,Carvunis:2021jga}.

\subsection{Exclusive $B \to K^{(*)} \ell \ell$ and $B_s \to \phi \ell \ell$ decays} \label{sec:th:exclusive}

The rare semileptonic $b \to s \ell \ell$ decays $B \to K^{(*)} \mu \mu$ and $B_s \to \phi \mu \mu$ are sensitive to a large set of Wilson coefficients, most importantly $C_7^{(\prime)}$, $C_9^{(\prime)}$, and $C_{10}^{(\prime)}$. The relative sensitivity to the different Wilson coefficients varies across the available range of di-lepton invariant mass $q^2$. 
In addition to the branching ratio, the angular distribution of the decays provides various additional observables. In the case of the $B \to K \mu \mu$ decay there is the muon forward backward asymmetry $A_\text{FB}$ and the so-called flat term $F_H$~\cite{Bobeth:2007dw}. In the case of the four-body decays $B \to K^* (\to K \pi) \mu \mu$ and $B_s \to \phi (\to K K) \mu \mu$ one has access to many independent angular coefficients~\cite{Bobeth:2008ij, Altmannshofer:2008dz, Matias:2012xw}. Frequently, one works with so-called ``optimized observables'', combinations of angular coefficients which have reduced dependence on form factors~\cite{Matias:2012xw, Descotes-Genon:2012isb}. The most prominent of them is the optimized observable $P_5^\prime$ in the $B \to K^* (\to K \pi) \mu \mu$ decay.

The $B \to K^{(*)} \mu \mu$ and $B_s \to \phi \mu \mu$ observables can be expressed in terms of transversity amplitudes that depend on ``local'' and ``non-local'' hadronic matrix elements which are subject to sizable theory uncertainties~\cite{Jager:2012uw, Jager:2014rwa}. The local matrix elements are the $B \to K^{(*)}$ and $B_s \to \phi$ form factors. For small hadronic recoil, i.e. in the large $q^2$ region above the narrow charmonium resonances, the form factors can be determined from lattice QCD~\cite{Horgan:2013hoa, Bouchard:2013eph, Horgan:2015vla, Bailey:2015dka}. The existing lattice calculations of the form factors have uncertainties of better than $10\%$ and are expected to become more precise in the coming years~\cite{USQCD:2019hyg, Boyle:2022uba}. 
At low $q^2$, QCD light-cone sum rules (LCSR) have been used to obtain the form factors with an accuracy of $\sim 10\%$~\cite{Bharucha:2015bzk, Khodjamirian:2017fxg, Gubernari:2018wyi, Gao:2019lta}. The LCSR and lattice determinations can be extrapolated to the full $q^2$ range by parameterizing the $q^2$ dependence using the $z$-expansion. Comparing LCSR and lattice results, one finds reasonable agreement, and combined fits of the form factors exist~\cite{Bharucha:2015bzk, Gubernari:2018wyi}. The $B \to K^*$ and $B_s \to \phi$ form factors are calculated in the narrow width limit. Studies indicate that going beyond this limit for the $K^*$ gives $\mathcal O(10\%)$ corrections~\cite{Descotes-Genon:2019bud}. Working directly with $B \to K \pi$ form factors will become more and more important in the future. First lattice calculations are under way~\cite{USQCD:2019hyg, Boyle:2022uba}.

The non-local effects are much more challenging to estimate. The most relevant contributions to the non-local matrix elements are from ``charm loops'' that involve current-current four-quark operators. At low (or negative) $q^2$, a light-cone operator product expansion can be used to estimate these contributions. Existing calculations suggest that they are small~\cite{Khodjamirian:2010vf, Gubernari:2020eft}. However, given the observed anomalies in the exclusive decays, more scrutiny is needed. 
At high $q^2$, quark hadron duality is used to argue that the effect of the broad charmonium resonances averages out in sufficiently large $q^2$ bins. Attempts to describe the high $q^2$ spectrum locally are currently based on a naive factorization approach~\cite{Lyon:2014hpa, Brass:2016efg}. Assigning robust uncertainties to the non-local effects is difficult. 

Already the existing experimental measurements of the $B \to K^{(*)} \mu \mu$ and $B_s \to \phi \mu \mu$ branching ratios have reached a precision that surpasses the precision of the SM theory predictions. In the case of angular observables, the current experimental and theory uncertainties are comparable. To benefit from the expected experimental improvements it is therefore crucial to obtain much better control on the theory uncertainties. In this respect, data driven methods to determine the non-local effects will become more and more popular. Various parameterizations of the non-local contributions have been suggested~\cite{Bobeth:2017vxj, Blake:2017fyh, Chrzaszcz:2018yza, Gubernari:2020eft}. With sufficient data, it should be possible to determine the model parameters and possible local new physics contributions simultaneously in the context of global fits. If the parameterizations are generic enough, the information one obtains on the new physics is robust.
Ways to determine non-local effects from lattice QCD are also explored~\cite{Nakayama:2019eth, Boyle:2022uba}.

In the future, one expect precise experimental results also for exclusive $b \to d \ell \ell$ decays, e.g. $B \to \pi \mu\mu$, $B \to \rho \mu\mu$, and $B_s \to K^* \mu\mu$. In such decays, non-local effects are qualitatively different as contributions from $\rho$ and $\omega$ resonances are not suppressed~\cite{Hambrock:2015wka,Khodjamirian:2017fxg}. Controlling those effects will be crucial to establish exclusive $b \to d \ell \ell$ decays as important probes of new physics~\cite{Rusov:2019ixr}.

\subsection{Baryonic decays $\Lambda_b \to \Lambda \ell \ell$} \label{sec:th:baryons}

The baryonic decays $\Lambda_b \to \Lambda \ell\ell$ are also sensitive probes of new physics in $b\to s \ell \ell$ transitions. $\Lambda_b$ baryons are produced with a high rate at the LHC, and in recent years progress has been made in the theory description of these decays. Similarly to the exclusive $B$ meson decays, the $\Lambda_b \to \Lambda \mu \mu$ decay offers a multitude of observables, including branching ratios, angular distributions, and CP asymmetries~\cite{Boer:2014kda, Blake:2017une, Roy:2017dum, Das:2018sms, Yan:2019tgn}. Light cone sum rule and lattice QCD calculations exist for from factors for the $\Lambda_b \to \Lambda$ decays~\cite{Wang:2015ndk, Detmold:2016pkz, Boyle:2022uba}.

Baryonic decays are highly complementary to the meson decays. Combining the results from the pseudoscalar to pseudoscalar transitions $B \to K \mu\mu$, the pseudoscalar to vector transitions $B\to K^* \mu\mu$ and $B_s \to \phi \mu\mu$, and the fermion to fermion transitions $\Lambda_b \to \Lambda \mu \mu$, is a powerful way to probe physics beyond the Standard Model.
Wilson coefficient fits that take into account the existing experimental results on the $\Lambda_b \to \Lambda (\to p \pi) \mu \mu$ angular distribution and the $\Lambda \to p \pi$ decay asymmetry parameter find relevant constraints on the coefficients $C_9$ and $C_{10}$~\cite{Meinel:2016grj,Blake:2019guk}.
Important consistency checks for theory computations and experiment are provided by kinematic endpoint relations~\cite{Hiller:2021zth}.

Interesting complementary information on new physics in $b \to s \mu\mu$ transitions might also come from decays to excited $\Lambda$ baryons, in particular $\Lambda_b \to \Lambda(1520) (\to p K) \mu \mu$~\cite{Descotes-Genon:2019dbw, Das:2020cpv, Amhis:2020phx}. First lattice calculations of the relevant $\Lambda_b \to \Lambda(1520)$ form factors already exist~\cite{Meinel:2020owd, Meinel:2021mdj, Boyle:2022uba}.

\subsection{Inclusive $B \to X_s \gamma$ and $B \to X_s \ell \ell$ decays} \label{sec:th:inclusive}

The inclusive $B \to X_s \gamma$ decay is known to be a highly sensitive probe of new physics. In the language of the effective Hamiltonian, the $B \to X_s \gamma$ decay gives one of the tightest constraints on new physics contributions to the Wilson coefficients $C_7$ and $C_7^\prime$~\cite{Paul:2016urs}.  
The measured CP- and isospin-averaged $B \to X_s \gamma$ branching ratio agrees well with the corresponding SM prediction~\cite{Amhis:2019ckw, Misiak:2020vlo}
\begin{equation}
 \text{BR}(B \to X_s \gamma)_\text{exp} = (3.32 \pm 0.15) \times 10^{-4} ~,\quad \text{BR}(B \to X_s \gamma)_\text{SM} = (3.40 \pm 0.17) \times 10^{-4} ~.
\end{equation}
The experimental and theoretical values have reached a remarkable precision of $\lesssim 5\%$. At Belle II, an improvement of the experimental uncertainty by approximately a factor of $\sim 2$ can be expected~\cite{Belle-II:2018jsg}. Achieving a similar precision on the theory side is challenging but essential to improve the constraining power of the $B \to X_s \gamma$ decay. It requires the evaluation of the Next-to-Next-to Leading (NNLO) QCD corrections without interpolation in the charm mass, as well as control over non-perturbative effects that are expected to give few $\%$ effects~\cite{Benzke:2010js, Gunawardana:2019gep, Benzke:2020htm}. 

The branching ratios quoted above are given for a cut on the photon energy of $E_\gamma > 1.6$\,GeV. Already at the current level of precision and even more so in the future, it is very important to control uncertainties related to this cut. Alternatively, one can use fits to the $E_\gamma$ spectrum to constrain $C_7^{(\prime)}$ more robustly~\cite{Bernlochner:2020jlt}.

\bigskip
The inclusive $B \to X_s \ell \ell$ decays are very important probes of new physics~\cite{Huber:2020vup}. In contrast to the $B \to X_s \gamma$ decay, they give access not only to $C_7^{(\prime)}$ but also to the semileptonic Wilson coefficients $C_9^{(\prime)}$ and $C_{10}^{(\prime)}$. Compared to the exclusive decays discussed in section~\ref{sec:th:exclusive} the hadronic uncertainties are under much better theoretical control, in particular at low $q^2$ (i.e. for a di-lepton invariant mass below the charmonium resonances). In addition to the $B \to X_s \ell \ell$ branching ratio, an angular analysis in the angle between the positively charged lepton and the initial $B$ meson in the di-lepton center of mass frame gives access to two more observables: the forward backward asymmetry, $A_\text{FB}$, and the equivalent to the longitudinal polarization fraction in exclusive decays, $F_L$~\cite{Lee:2006gs}.

At low $q^2$, the existing theory predictions include NNLO QCD and NLO electroweak corrections as well as the leading power corrections to the heavy quark limit. At high $q^2$, the heavy mass expansion breaks down at the kinematic endpoint, but power corrections can be largely tamed by normalizing to the semileptonic $B \to X_u \ell \nu$ decay rate with the same $q^2$ cut~\cite{Ligeti:2007sn}. QED corrections to the differential branching ratio are also important and care must be taken to use the same prescription for QED radiation when comparing theory predictions to experimental results~\cite{Huber:2015sra}. 
Overall, the SM predictions for $B \to X_s \ell \ell$ have reached a precision of $\sim 6\%$ at low $q^2$ and of $\sim 12\%$ at high $q^2$~\cite{Huber:2020vup}. 

It is expected that Belle II will be able to measure the $B \to X_s \ell \ell$ rates with a precision of $\lesssim 5\%$~\cite{Belle-II:2018jsg}, which calls for further improvements of the SM predictions. This requires in particular studying the impact of the cut on the hadronic invariant mass.

\subsection{Di-neutrino decays $b \to s \nu\bar\nu$} \label{sec:th:dineutrino}

The rare decays $B \to K \nu\bar\nu$ and $B \to K^* \nu\bar\nu$ are well known to be sensitive probes of BSM physics~\cite{Altmannshofer:2009ma, Buras:2014fpa, Browder:2021hbl, Bause:2021cna}. 
Complementary information can come from the related modes $B_s \to \phi\nu\bar\nu$, $\Lambda_b \to \Lambda\nu\bar\nu$, and $B_c \to D_s \nu\bar\nu$. Combining the information from the whole family of di-neutrino modes, i.e. the pseudoscalar to pseudoscalar
transitions $B \to K \nu\bar\nu$ and $B_c \to D_s \nu\bar\nu$, the pseudoscalar to vector transitions $B \to K^* \nu\bar\nu$
and $B_s \to \phi\nu\bar\nu$, and the fermion to fermion transition $\Lambda_b \to \Lambda\nu\bar\nu$, can be a powerful way to probe new physics.
The $B \to K^{(*)} \nu\bar\nu$ decays will be measured with good precision at Belle II, while the $B_s$, $B_c$ and $\Lambda_b$ modes could be accessed at the $Z$-pole runs of the proposed FCC-ee and CEPC.

From the theory point of view, the di-neutrino modes are cleaner than the corresponding exclusive decays with charged leptons based on the $b \to s \ell \ell$ transition. In fact, the di-neutrino modes are not affected by the non-local hadronic contributions (aka charm loops), that are a major source of theory uncertainty in decays like $B \to K^{(*)} \ell^- \ell^+$ and $B_s \to \phi \ell^+ \ell^-$. All hadronic physics in the $b \to s \nu\bar\nu$ decays can be captured by local form factors. As discussed above in sections~\ref{sec:th:exclusive} and~\ref{sec:th:baryons}, the relevant form factors are calculated with light-cone sum rule and lattice QCD methods with $O(10\%)$ precision. Improvements from the lattice are expected on the relevant time-scales~\cite{Boyle:2022uba}.

The role of the di-neutrino modes in testing new physics is two-fold. (i) they probe heavy new physics that can be model independently described by 4 fermion contact interactions. Due to the $SU(2)_L$ gauge symmetry, one generically expects that new physics effects in $b \to s \nu\bar\nu$ and $b \to s \ell^+\ell^-$ decays are related and give complementary information about the new physics. As the neutrino flavor is not observed in the experiments, the di-neutrino modes can give important indirect information about $b \to s \tau^+\tau^-$ decays that are currently only weakly constrained. (ii) the measurements of the di-neutrino decays might also give access to light dark sectors (see e.g.~\cite{MartinCamalich:2020dfe, Hostert:2020gou, Felkl:2021uxi, Crivellin:2022obd} for recent studies). The decays $b \to s X$, where $X$ corresponds to one or more invisibly decaying or neutral long lived light new particle gives the same missing energy signature as the di-neutrino decays. Examples of such light particles include dark photons, light $Z^\prime$ gauge bosons, sterile neutrinos, axions, or neutral scalars. The presence of the light new particles in the decay can significantly change the kinematical distributions and dedicated experimental studies might be required to consistently probe light new physics scenarios.

\subsection{The future of global fits of rare $b$ hadron decays} \label{sec:th:fits}

To fully exploit the complementarity of the vast set of rare decay obserables to new physics, global fits to all relevant rare $b$ decay data are often performed. The outcome of those fits are constraints or best fit regions for the new physics Wilson coefficients of the effective Hamiltonian in equation~\eqref{eq:Heff}. The global fits have reached a high level of sophistication. They take into account measurements of more than 100 observables in $B$ meson and $\Lambda_b$ baryon decays and incorporate correlations of the experimental and theoretical uncertainties.

Various groups are performing global fits of the data, including different sets of observables, employing different treatments of hadronic uncertainties, and making use of different statistical methods. Many publicly available tools~\cite{Straub:2018kue, Aebischer:2018bkb, Aebischer:2018iyb, DeBlas:2019ehy, Neshatpour:2021nbn, vanDyk:2021sup, Sibidanov:2022gvb} can be used in global fits. Overall, there is a broad consensus that a new physics effect in the form of the four fermion contact interactions $C_9 (\bar s \gamma_\alpha P_L b)(\bar\mu \gamma^\alpha \mu)$ and $C_{10} (\bar s \gamma_\alpha P_L b)(\bar\mu \gamma^\alpha \gamma_5 \mu)$ gives a consistent description of all the currently observed $b \to s \ell \ell$ anomalies~\cite{Geng:2021nhg, Altmannshofer:2021qrr, Alguero:2021anc, Hurth:2021nsi, Isidori:2021vtc, Ciuchini:2021smi}. However, apart from the theoretically clean LFU test and the $B_s \to \mu^+\mu^-$ branching ratio, the relevant observables are affected by poorly known hadronic contributions, and the results of the fits have to be interpreted with caution.

Indeed, the treatment of hadronic uncertainties is one of the most critical aspects of global fits. In exclusive $b$ decays, the main uncertainties in the theoretical predictions come from the hadronic form factors, but also from non-factorizable effects (see the discussion in subsection~\ref{sec:th:exclusive} above). Assuming that the uncertainties are estimated in a sufficiently conservative manner, global fits find consistently strong evidence for new physics with pulls exceeding $5\sigma$. However, it is conceivable that unaccounted for hadronic effects could mimic lepton universal new physics.
In the future, it might be possible to extract hadronic effects directly from data in a global fit. Proposals in this direction~\cite{Bobeth:2017vxj, Blake:2017fyh, Chrzaszcz:2018yza, Gubernari:2020eft} provide the basis for such studies.
As long as the hadronic contributions are parameterized in a sufficiently generic way, any missing standard physics can be accounted for. It is expected that the large amount of data from LHCb will allow one to determine the hadronic effects directly from data and to simultaneously retain sensitivity to new physics.

In this respect, it is important to note that Belle II will soon be able to make precision measurements of the inclusive processes $B \to X_s \ell\ell$. As those decays are under good theoretical control at low di-lepton invariant mass (see subsection~\ref{sec:th:inclusive} above), they will be invaluable input to global fits to scrutinize the anomalies observed by LHCb in the exclusive $b\to s \ell \ell$ counterparts.

As discussed in section~\ref{sec:intro} one can make the argument that the rare decays based on the $b \to d \ell \ell$ transition, like $B\to \pi \mu\mu$ or $B_s \to K^* \mu\mu$, have even better sensitivity to new physics than $b \to s \ell \ell$ decays, as they are stronger suppressed in the SM. At the very least, the rare $b \to d \ell \ell$ decays give complementary information as they probe different new physics flavor structures. It is thus highly motivated to analyse these decay modes for new physics effects.
The branching ratios of $b \to d \ell \ell$ decays are at the level of $10^{-8}$ or smaller, making them a challenging target for experiment. Nevertheless, LHCb already reported evidence for e.g. the $B_s \to K^* \mu\mu$ decay~\cite{LHCb:2018rym}.
During the next runs of the LHC, the LHCb experiment will collect a sufficient amount of data which should allow them to measure a large set of interesting observables in the $b \to d \ell \ell$ decays, like angular observables in $B_s \to K^* \mu\mu$ or the LFU ratio BR$(B\to \pi \mu\mu)/$BR$(B\to \pi ee)$.
This calls for an extension of global fit framework including a careful assessment of the long-distance hadronic effects in $b \to d \ell \ell$ decays.

In the long term, future circular electron-positron colliders like CEPC and FCC-ee~\cite{CEPCStudyGroup:2018ghi, FCC:2018byv, Bernardi:2022hny, Cheng:2022zyy} offer unique sensitivities to rare decays that are not accessible at the LHC or Belle II. Running on the $Z$ pole could produce up to $10^{12}$ $b$ hadrons with a large boost in a clean environment. Among the most interesting processes that can be accessed at such experiments are rare $b$ hadron decays with taus in the final state like $B \to K^* \tau \tau$~\cite{Kamenik:2017ghi, Li:2020bvr, Bernardi:2022hny}.
Results on the tau modes would complete the picture of leptonic and semi-leptonic rare decays.

\subsection{Rare charm decays}

Compared to rare decays of $b$ hadrons, rare charm decays are much less explored so far. Due to the strong GIM suppression, the sensitivity of rare charm decays to new physics is in principle even higher than the one of rare $b$ decays~\cite{Burdman:2001tf,Fajfer:2015mia,deBoer:2015boa,Bharucha:2020eup}. However, in practice it is much more challenging to control the hadronic physics in charm decays. In particular, rare charm decays are subject to large resonance contributions that can dominate the short-distance contributions by orders of magnitude. Moreover, heavy quark methods are much less reliable compared to $b$ decays. This prohibits one to probe short-distance physics (be it from the SM or from new physics) in simple observables such as branching ratios. A considerable focus has therefore been on so-called ``null tests'', i.e. observables that are strongly suppressed in the SM due to exact or approximate symmetries and largely free of hadronic uncertainties. Observation of a non-standard effect in such null tests would be robust evidence for new physics. The new physics that could lead to non-standard effects in rare charm decays can also be probed through measurements of di-lepton production at the LHC~\cite{Fuentes-Martin:2020lea}. Examples of null tests include deviations from lepton flavor universality in semileptonic rare decays, like $D \to \pi \ell\ell$ and $D_s \to K \ell \ell$ and lepton flavor violating decays, like $D \to \pi e \mu$ and $D_s \to K e \mu$~\cite{Bause:2019vpr}. Such processes are covered in~\cite{Guadagnoli:2022oxk}. Here, we discuss additional null test, namely angular observables in semileptonic decays and di-neutrino decay modes, like $D \to \pi \nu\bar\nu$ and $D_s \to K \nu\bar\nu$~\cite{Bause:2020xzj}

The full angular distribution in the decays $D \to P_1 P_2 \ell \ell$, where $P_1$ and $P_2$ are a pion or Kaon, provides several null tests of the SM~\cite{DeBoer:2018pdx}. For example, the forward backward asymmetry $A_\text{FB}$ is proportional to the Wilson coefficient $C_{10}$, which corresponds to a leptonic axial-vector current. In the SM, the axial-vector in rare charm decays receives short distance contributions that show an exceptionally strong GIM suppression and long distance contributions arise only at higher order in QED. Rare charm decay observables that are proportional to $C_{10}$ are therefore vanishingly small in the SM. A non-zero observation of a forward backward asymmetry in $D \to P_1 P_2 \ell \ell$ would be a clear sign of new physics. Similarly, also several angular CP asymmetries in $D \to P_1 P_2 \ell \ell$ are predicted to be negligibly small in the SM~\cite{DeBoer:2018pdx}.
Also angular distributions in 3-body and 4-body rare charm baryon decays, like $\Lambda_c \to p \ell \ell$ or $\Xi_c \to \Lambda (\to p \pi) \ell \ell$, provide many angular observables that can serve as null test and that have complementary sensitivity to new physics~\cite{Golz:2021imq,Golz:2022alh}.

Furthermore, CP asymmetries and the photon polarization in radiative charm decays can be used to test the SM. For example, the SM predictions for the direct CP asymmetries in pure weak annihilation modes like $D^0 \to (K^{*},\bar K^*) \gamma$, or $D_s \to \rho \gamma$ vanish, and the CP asymmetries in $D \to \rho \gamma$ are expected to be very small~\cite{deBoer:2017que}. Also CP asymmetries in radiative 3-body charm meson decays are sensitive probes of new physics~\cite{Adolph:2020ema}. An untagged time-dependent analysis of $D^0$ decays into CP eigenstates would allow one to measure the photon polarization in the radiative decays~\cite{deBoer:2018zhz}. In the SM, the photon polarization can be estimated using $U$-spin symmetry. The photon polarization can also be determined through the up-down asymmetry in $D \to K_1 (\to K \pi\pi) \gamma$~\cite{Adolph:2018hde}.
Finally, the measurement of the photon polarization in radiative decays of charmed baryons has also been considered as a test of new physics~\cite{Adolph:2022ujd}.

Another class of null-tests are the di-neutrino transitions $c \to u \nu \bar\nu$. They are strongly GIM-suppressed in the SM and any observation with current and expected experimental sensitivities~\cite{BESIII:2021slf} would be a clean signal of new physics~\cite{Bause:2020xzj,Faisel:2020php}. Searches for the di-neutrino modes can also be sensitive to invisible light new physics particles like sterile neutrinos or dark photons that can be produced in charm decays~\cite{Su:2020yze}.

We note that null tests might provide robust evidence for a non-standard effect, but interpreting constraints, or a possible signal, in a new physics scenario requires the knowledge of hadronic parameters. Obtaining a better understanding of the long distance physics that governs rare charm decays therefore remains an interesting and important endeavor.

\section{Conclusions}

Rare decays of $b$ and $c$ flavored hadrons will continue to play a very important role in the search for physics beyond the Standard Model for many years to come. Rare decays have very high indirect sensitivity to new physics at very high scales. 
A comprehensive coverage of new physics parameter space requires the study of a large set of complementary rare decays.

In this whitepaper, we reviewed the current experimental status of the rare $b$ and $c$ decay program, and highlighted some of the opportunities at Belle II, BES III, the LHC (ATLAS, CMS, and LHCb) as well as proposed future particle colliders. To fully profit from the expected experimental results one requires an improved theoretical description of rare decays as well. We therefore also summarized some of the most important theory challenges that need to be tackled in the coming years. The expected experimental precision combined with improved theoretical predictions will allow us to probe uncharted new physics parameter space.

The current ``$B$ anomalies'' might be first indirect signs of new physics. If these hints were to be confirmed, they would imply a new mass scale in particle physics, potentially within reach of either the LHC or a new generation of colliders. Conversely, in the absence of anomalies, the described future rare decay program will quantitatively and qualitatively improve constraints on new physics and provide critical input for new physics model building.

\section*{Acknowledgements}
The research of W.A. is supported by the U.S. Department of Energy grant number DE-SC0010107. W.A. also acknowledges support by the Munich Institute for Astro- and Particle Physics (MIAPP) which is funded by the Deutsche Forschungsgemeinschaft (DFG, German Research Foundation) under Germany's Excellence Strategy – EXC-2094-390783311.